\documentclass[fleqn,usenatbib]{mnras}

\usepackage{newtxtext,newtxmath}

\usepackage[T1]{fontenc}

\DeclareRobustCommand{\VAN}[3]{#2}
\let\VANthebibliography\thebibliography
\def\thebibliography{\DeclareRobustCommand{\VAN}[3]{##3}\VANthebibliography}

\usepackage{ae,aecompl}
\usepackage{graphicx}	

\usepackage{ulem}
\usepackage{multirow}
\newcommand{\angstrom}{\textup{\AA}}

\newcommand{\UM}{\textsc{UniverseMachine}}

\newcommand{\Msun}{\mathrm{M}_\odot}
\newcommand{\appropto}{\mathrel{\vcenter{
  \offinterlineskip\halign{\hfil$##$\cr
    \propto\cr\noalign{\kern2pt}\sim\cr\noalign{\kern-2pt}}}}}

\title[Testing DM-Galaxy Correlation Strengths]{Observing Correlations Between Dark Matter Accretion and Galaxy Growth: II. Testing the Impact of Galaxy Mass, Star Formation Indicator, and Neighbour Colours}

\author[C. O'Donnell et al.]{
Christine O'Donnell$^{1}$\thanks{E-mail: Christine.ODonnell@asu.edu (CO)},
Peter Behroozi$^{2}$,
Surhud More$^{3,4}$
\\
$^{1}$School of Earth \& Space Exploration, Arizona State University, Tempe, AZ 85281, USA\\
$^{2}$Department of Astronomy and Steward Observatory, University of Arizona, Tucson, AZ 85721, USA\\
$^{3}$Inter-University Centre for Astronomy and Astrophysics, Post bag 4, Ganeshkhind, Pune 411007, India\\
$^{4}$Kavli Institute for the Physics and Mathematics of the Universe, 5-1-5 Kashiwanoha, Kashiwa, 2778583, Japan\\
}

\date{Accepted XXX. Received YYY; in original form ZZZ}

\pubyear{2021}

\begin{document}
\label{firstpage}
\pagerange{\pageref{firstpage}--\pageref{lastpage}}
\maketitle

\begin{abstract}
A crucial question in galaxy formation is what role new accretion has in star formation. Theoretical models have predicted a wide range of correlation strengths between halo accretion and galaxy star formation. Previously, we presented a technique to observationally constrain this correlation strength for isolated Milky Way-mass galaxies at $z\sim 0.12$, based on the correlation between halo accretion and the density profile of neighbouring galaxies. By applying this technique to both observational data from the Sloan Digital Sky Survey and simulation data from the \textsc{UniverseMachine}, where we can test different correlation strengths, we ruled out positive correlations between dark matter accretion and recent star formation activity. In this work, we expand our analysis by (1) applying our technique separately to red and blue neighbouring galaxies, which trace different infall populations, (2) correlating dark matter accretion rates with $D_{n}4000$ measurements as a longer-term quiescence indicator than instantaneous star-formation rates, and (3) analyzing higher-mass isolated central galaxies with $10^{11.0} < M_*/M_\odot < 10^{11.5}$ out to $z\sim 0.18$. In all cases, our results are consistent with non-positive correlation strengths with $\gtrsim 85$ per cent confidence, which is most consistent with models where processes such as gas recycling dominate star formation in massive $z=0$ galaxies. 
\end{abstract}




\section{Introduction}
\label{sec:intro}

According to the $\Lambda$CDM framework, galaxies form within dark matter haloes when gas gravitationally coalesces at halo centres \citep[for reviews, see][]{Somervile15,WT18}. Thus, we expect that halo properties and galaxy properties will be strongly correlated, e.g., halo mass and stellar mass \citep{Tinker17b, Behroozi19}.

However, different models predict different correlation strengths between dark matter accretion and galaxy star formation. As material falls onto a halo from large distances, we expect the fraction of infalling gas versus infalling dark matter to match the cosmic baryon fraction. If this also holds true at smaller scales, then we would expect dark matter accretion and star formation to be correlated. For example, \cite{WetzelNagai15} found that dark matter accretes in a shell-like manner at $R_{200\mathrm{m}}$ around a halo. Gas, on the other hand, can radiatively cool, allowing it to decouple from the dark matter and continue infalling onto the central galaxy. As a result, star formation rates track dark matter accretion rates \citep{WetzelNagai15}, and many theoretical models and simulations have found or assumed a perfect positive correlation strength between the two \citep[e.g.,][]{Becker15, Rodriguez-Puebla16, Cohn17, Moster18}.

On the other hand, some models predict that feedback from winds, supernovae, AGN, and other processes will suppress new accretion onto central galaxies. Thus, most star formation is generated by recycled or re-accreted gas, and we would expect at most only a weak correlation with dark matter accretion \citep[e.g.,][]{Keres05, Dekel06, Nelson13, Nelson15, Muratov15, vandeVoort16}. Furthermore, \cite{Muratov15} found that outflows from a galaxy (due to supernovae, AGN, etc.) are most significant at higher redshifts, creating an enriched gas reservoir that powers star formation at lower redshifts. These models are consistent with observational results that star formation rates do not correlate with major mergers \citep{Behroozi15}, which have enhanced dark matter accretion rates. Further, \cite{Tinker17} studied SDSS galaxy groups and found that the fraction of quenched central galaxies with $M_* \gtrsim 10^{10.3} \Msun$ only slightly increases as the local environmental density increases. However, halo assembly rates are strongly correlated with local density \citep[e.g.,][]{Lee17}, and so their results implied that halo growth and galaxy assembly are only weakly correlated. 

In \cite{ODonnell20}, we observationally constrained the correlation between dark matter accretion and recent star formation activity in Milky-Way mass galaxies ($10^{10.5} < M_\ast / \Msun < 10^{11}$). Our technique built on work to characterize the splashback radii of haloes, the radius at which newly accreted material reaches its first apocentre \citep[e.g., ][]{Diemer14, More15, More16, Baxter17}. s a halo accretes more matter, its gravitational potential well deepens, which will tighten the orbits of satellite galaxies and thus steepen the halo's density profile. For more rapidly accreting haloes, their halo density profile steepens more strongly \citep{Diemer14} and their splashback radii will decrease \citep{More15}.
Observational studies stacked the density profiles of nearby neighbours around thousands of clusters to look for the splashback feature by measuring excess galaxy counts around target clusters using background subtraction on photometric SDSS data \citep{More15, More16, Baxter17}.

In our previous analysis, we also used neighbouring galaxies as probes of dark matter accretion.  To measure correlations with galaxy star formation rates, we made two modifications to previous techniques. First, we selected Milky Way-mass galaxies, as star formation is still happening at these smaller mass scales (versus the centrals of galaxy clusters, which are often quenched). To reduce environmental contamination in neighbour density profiles, we specifically selected \textit{isolated} Milky Way-mass galaxies.  By selecting isolated galaxies that are the dominant source of gravity in their local environments, they will have stronger correlations between neighbouring galaxy orbits and dark matter accretion rates \citep[see also][]{Deason20}, which allows us to probe lower-mass halo scales than previous work. In addition, instead of identifying a single feature in the density profiles, we analyzed the shape of the entire neighbour density distribution to increase our signal-to-noise ratio. By comparing the measured shapes of the neighbour density distributions, our technique allows us to assess the dark matter accretion rates.

Our analysis compared observational SDSS DR16 data \citep{SDSS_DR16} to simulated \UM{} data \citep{Behroozi19} to constrain the correlation strength. We separated star-forming and quiescent isolated galaxies in the SDSS based on their specific star formation rates (SSFRs). Our results ruled out positive correlations between dark matter accretion rates and SSFRs with $\gtrsim 85$ per cent confidence.

This paper extends our previous work by ruling out several alternate interpretations of this finding.  For example, we would expect weak correlations if the timescales probed by SSFRs are much shorter than the orbits of satellite galaxies ($\sim 2 t_\mathrm{dyn} \sim 4$ Gyr; see \S{}5 in \citealt{ODonnell20}). In this paper, we test two approaches that address this concern:
\begin{enumerate}
    \item Instead of only separating star-forming and quiescent galaxies based on their SSFRs, we also bin galaxies based on their 4000\AA{} break \citep[$D_{n}4000$; ][]{Balogh99}, which is a longer-term quiescence indicator.
    \item We compare the density distributions of neighbouring galaxies based on the neighbours' colours. As a satellite galaxy falls into a host halo, gas is stripped from the satellite that would otherwise replenish star formation, leading to an increase in the fraction of red galaxies within host halo virial radii \citep[e.g., ][]{Gunn72, Moore96, Dressler97, Weinmann06, Kawata08, Baxter17}. \cite{Wetzel13} found that the typical timescale for this quenching is on the order of satellite orbital periods (2-4 Gyr). Because red satellites have been within their host haloes for a longer time, they may be more sensitive to changes in the gravitational potential well than blue satellites that have only recently fallen in. By analyzing the density distribution of red neighbours around isolated Milky Way-mass galaxies, we would have a more robust test of the correlation strength between dark matter accretion and star formation rates. 
\end{enumerate}
Furthermore, we expand our analysis to higher-mass isolated host galaxies. This test allows us to identify isolated central galaxies out to higher redshifts (up to $z<0.183$ versus $z<0.123$; see \S 2.1.1 for sample statistics), and it adds an additional check of our results by using an independent host population.

This paper is structured as follows: First, in \S\ref{sec:data}, we summarise key details of our observational (\S\ref{sec:obs_data}) and simulation data (\S\ref{sec:th_data}), including differences with the datasets used in \cite{ODonnell20}. In \S\ref{sec:methods}, we describe the methodology used in our analysis. \S\ref{sec:results} presents the results for separating star-forming and quiescent hosts based on SSFR versus $D_{n}4000$ (\S\ref{sec:results_sf}), comparing the density distributions of red neighbours around isolated hosts to the distributions of blue neighbours (\S\ref{sec:results_red}), and analysing higher-mass isolated hosts (\S\ref{sec:results_hostmass}). Finally, we conclude in \S\ref{sec:disc_conclu} and note directions for future analyses.
We adopt a flat $\Lambda$CDM cosmology with $\Omega_M = 0.307$,  $\Omega_{\Lambda} = 0.693$, and $h = 0.677$, consistent with \textit{Planck} 2018 results \citep{Planck18}

\section{Observations \& Simulations}
\label{sec:data}

This paper uses similar techniques and datasets as in \cite{ODonnell20}. Below, we repeat key details and note differences where appropriate.

\subsection{Observational Data}
\label{sec:obs_data}

We identify isolated galaxies, which we refer to as our \textit{isolated host} sample (\S\ref{sec:methods}), from the SDSS DR16 spectroscopic catalogs \citep{SDSS_DR16}. We define \textit{isolated} to mean that there is no larger galaxy within 2 Mpc in projected (on-sky) physical distance or 1000 km/s in velocity separation. We use median stellar masses, specific star formation rates, and $D_{n}4000$ values from the MPA-JHU value-added catalog \citep{Kauffmann03, Brinchmann04}. Stellar masses and star formation rates were converted to a \cite{Charbrier03} IMF by dividing each by a factor of 1.07. To improve our isolated host selection, we supplemented these catalogs with data from the NYU Value-Added Galaxy Catalog \citep[NYU-VAGC;][]{Blanton_VAGC} for galaxies with $M_* > 10^{9.5} M_\odot$. The NYU-VAGC filled in information for galaxies affected by fibre collisions by assuming they have the same redshift as the nearest non-fibre-collided neighbour. Further, we excluded galaxies that are within 2 Mpc of a survey boundary or region of significant incompleteness to ensure the robustness of our isolation criteria. To avoid Hubble flow corrections \citep[e.g.,][]{Baldry12}, we exclude galaxies with $z < 0.01$. Our resulting catalog has 547,271 galaxies over 6401.1 deg$^2$ of sky. Finally, we apply a stellar mass completeness cut to our spectroscopic catalog. \cite{Behroozi15} found that in the SDSS, $>95$ per cent of galaxies have $r$-band apparent magnitudes ($r$) brighter than the following limit:
\begin{equation}
    r < -0.25 - 1.9 \log_{10} \left( \frac{M_*}{M_\odot} \right) + 5 \log_{10} \left( \frac{D_L(z)}{10\mathrm{pc}} \right) \, ,
    \label{eq:SM_complete_cut}
\end{equation}
where $M_*$ is the stellar mass and $D_{L}$ is the luminosity distance given our cosmology. To be consistent with SDSS's spectroscopic survey limits, we exclude galaxies for which $r > 17.77$ according to Eq.~\ref{eq:SM_complete_cut}. 

In this paper, we identified isolated host galaxies in two different mass bins: (1) galaxies with $10.5 < \log_{10}(M_*/M_\odot) < 11.0$ and (2) galaxies with $11.0 < \log_{10}(M_*/M_\odot) < 11.5$; the former is the same bin we used in \cite{ODonnell20}.  

To count neighbouring photometric galaxies around our isolated hosts, we use SDSS DR16 photometric catalogs \citep{SDSS_DR16}. We use sources with a type field of `GALAXY' to exclude likely stars, and we restrict our catalog to galaxies with $r<21.5$ to ensure reliability of $g-r$ colours. Following \cite{ODonnell20}, we bin nearby neighbours by stellar mass to reliably compare the shape of density distributions around star-forming and quiescent hosts. We used the same fit between $g-r$ colours and $M_*/L_r$ ratios as found in \cite{ODonnell20}:
\begin{equation}
    \log_{10}(M_*/L_r) = 1.341 \, (g-r) - 0.639 \, .
    \label{eq:MsLr_gr}
\end{equation}
In \cite{ODonnell20}, we found that our results using this approach were consistent with those using luminosity binning \citep[Appendix A1 of][]{ODonnell20} and with using a fit between $g-r$ colours and mass from \cite{Bell03} \citep[Appendix A2 of][]{ODonnell20}.

\begin{figure*}
    \centering
    \includegraphics[width=\textwidth]{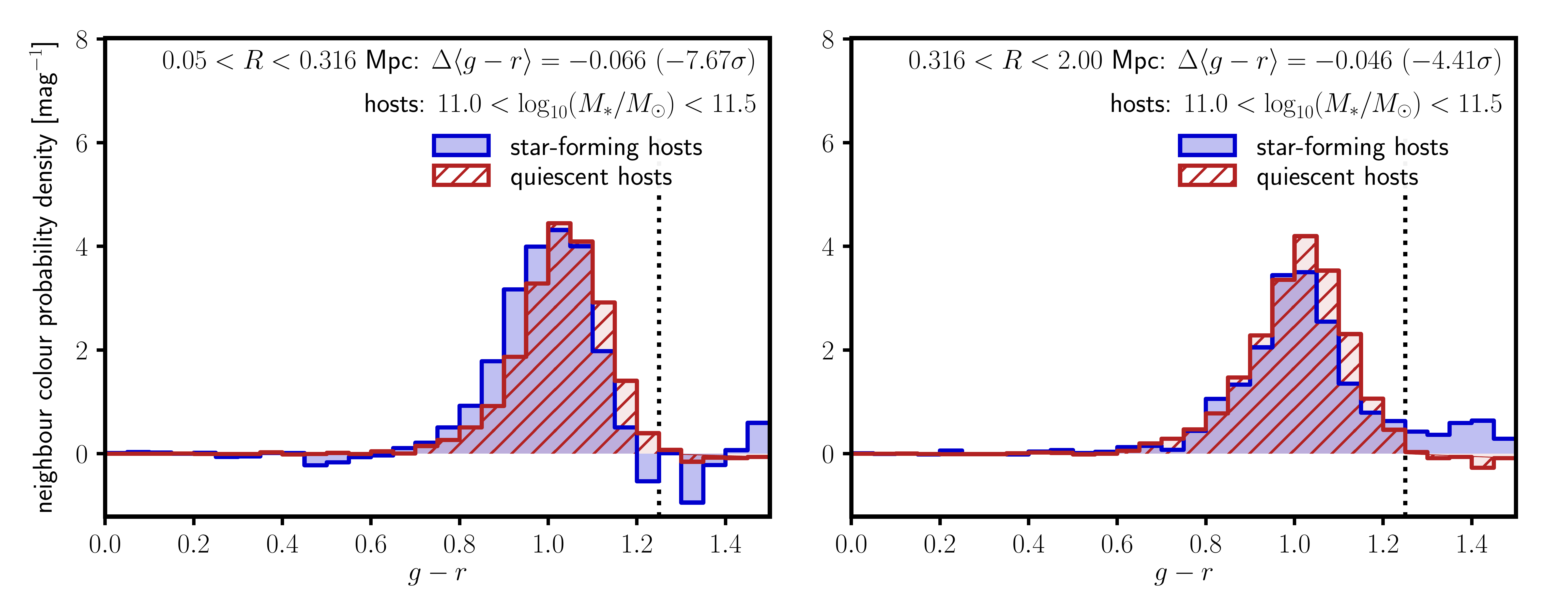}\\[-3ex]
    \caption{We exclude photometric galaxies with very red colours as they tend to be at higher redshifts \citep[e.g.,][]{redmapperI}; thus, applying a colour cut reduces noise in our neighbour density distributions. In \protect\cite{ODonnell20}, we excluded photometric galaxies with $g-r > 1.0$ for isolated hosts with $10.5 < \log_{10}(M_*/M_\odot) < 11.0$ as these galaxies were not present above background noise counts. Here, we plot the background-subtracted weighted distribution of $g-r$ colours for our higher-mass isolated hosts ($11.0 < \log_{10}(M_*/M_\odot) < 11.5$) and determine that the colour cut should be $g-r > 1.25$ (indicated by the dotted vertical line).  These plots include neighbours with $\log_{10}(M_*/M_\odot) > 10.4$, which corresponds to the stellar mass limit at the maximum isolated host redshift ($z=0.183$) given SDSS photometric limits. We note that lower-mass neighbours are expected to have bluer colours. The projected distance ranges of the two panels match the regions used in our analysis of the shapes of the neighbour density distributions (Eq.~\ref{eq:shape_ratio} in \S\ref{sec:methods}).  Neighbours around star-forming hosts have bluer $g-r$ colours than neighbours around quiescent hosts, and the difference is more significant at closer distances from the hosts. We noted a similar difference in the neighbours around isolated hosts with $10.5 < \log_{10}(M_*/M_\odot) < 11.0$ \citep[Fig. 9 in ][]{ODonnell20}.}
    \label{fig:neighbour_gr_highmass}
\end{figure*}

To reduce noise when applying our fit, we cut our photometric catalog based on $g-r$ colours to exclude galaxies at higher redshifts. For isolated hosts with $10.5 < \log_{10}(M_*/M_\odot) < 11.0$, we restrict our analysis to galaxies with $0.0 < g-r < 1.0$, as redder galaxies are not present above background noise levels \citep[Fig. 9 from][]{ODonnell20} and tend to be at higher redshifts \citep[e.g.,][]{redmapperI}. These cuts result in a photometric catalog that includes 35,457,243 galaxies over an on-sky area of 18,509.0 deg$^2$. We note that our results in \cite{ODonnell20} were not sensitive to the $g-r$ cutoff value; we obtained consistent results using a redder cutoff of $0.0 < g-r < 1.25$. 
The higher-mass isolated hosts ($11.0 < \log_{10}(M_*/M_\odot) < 11.5$) can be detected at higher redshifts, so we use a limit of $g-r < 1.25$ based on the colour distribution of nearby neighbours (Fig.~\ref{fig:neighbour_gr_highmass}), resulting in a photometric catalog that includes 47,713,412 galaxies.

\subsubsection{Sample Statistics}
\label{sec:obs_stats}

From \cite{ODonnell20}, we identified 25,625 isolated galaxies from SDSS with stellar masses $10.5 < \log_{10}(M_*/M_\odot) < 11.0$ that correspond to a redshift range of $0.01 < z < 0.123$ (median $z=0.079$). In this paper, we also identify 25,432 isolated hosts with stellar masses $11.0 < \log_{10}(M_*/M_\odot) < 11.5$ (redshift range $0.01 < z < 0.183$, with median $z=0.116$).\footnote{The isolated galaxy catalogs are available at \url{https://github.com/caodonnell/DM_accretion}} We also investigated using galaxies from a lower mass range ($10.0 < \log_{10}(M_*/M_\odot) < 10.5$), but their neighbour density distributions were dominated by noise because there were too few isolated hosts even if we relaxed the isolation criteria (e.g., no larger galaxy within 1 Mpc projected distance and 1000 km/s velocity distance).

To measure the uncertainties in neighbour density distributions, we used jackknife sampling. For each sample, a $\sim 10^\circ \times 10^\circ$ region was removed from the sky footprint ($\sim 37.5 \times 37.5$ Mpc/$h$ at $z=0.079$), resulting in 112 samples with an average of $\sim$25,000 isolated hosts in each mass bin per sample.

\subsubsection{Star Formation \& Quiescence Indicators}
\label{sec:sf_indicators}

In \cite{ODonnell20}, we binned our isolated hosts into star-forming and quiescent bins based on their specific star formation rates (SSFRs), which is an indicator of recent star formation activity. We separated the two SSFR bins at $\mathrm{SSFR} = 10^{-11} \mathrm{yr}^{-1}$ following \cite{Wetzel12}, and we keep the same definition here. As an additional test of our results, we attempted to separate the star-forming hosts into two bins since it is possible the correlation between star formation and dark matter accretion may differ for galaxies with stronger versus weaker star formation rates \citep{Berti12}. However, even when separating the star-forming hosts into two bins by the median SSFR, we did not have sufficient signal-to-noise to identify differences in their neighbour density distributions.

\begin{figure}
    \centering
    \includegraphics[width=\columnwidth]{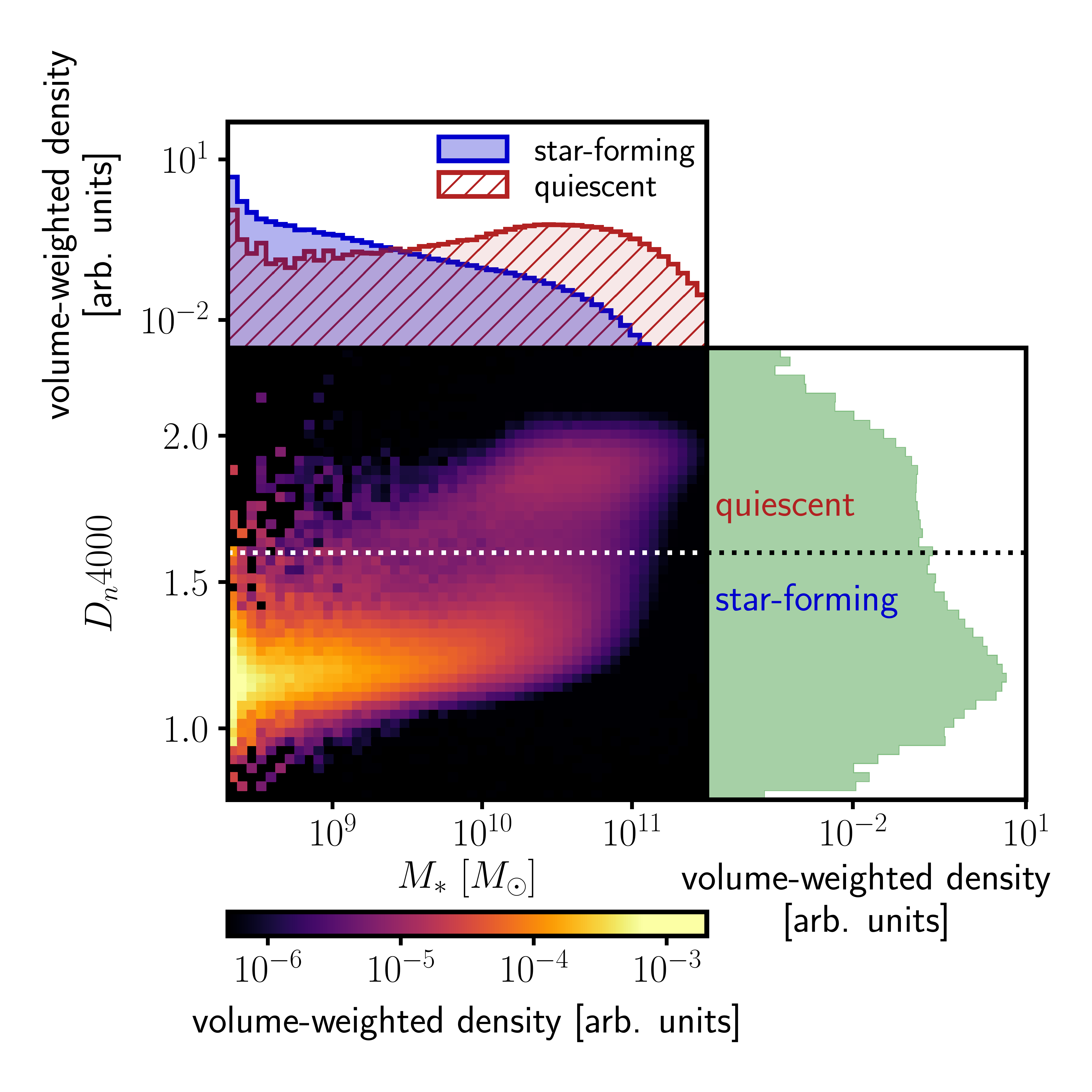}
    \caption{When using $D_{n}4000$ as a star formation indicator, we bin our isolated hosts into star-forming and quiescent hosts split at $D_{n}4000 = 1.6$. The central plot shows the volume-weighted density distribution of galaxies in the SDSS DR16 spectroscopic catalog. The top histogram shows the distribution of stellar masses of star-forming versus quiescent hosts based on their $D_{n}4000$ values, and the right histogram shows the overall distribution of $D_{n}4000$. Fig.~6 in \protect\cite{ODonnell20} depicts analogous distributions when using specific star formation rates (SSFRs) as the star formation indicator.}
    \label{fig:sm_dn4000}
\end{figure}

As we noted in \cite{ODonnell20}, the shape of the neighbour density distribution changes on timescales of satellite galaxy orbits $\sim 2 t_\mathrm{dyn} \sim 4$ Gyr. If SSFRs change on shorter timescales than satellite galaxy orbits, then we would expect to see weaker correlations. To test this potential bias, we also split the isolated hosts into two bins based on their 4000\angstrom{} break \citep[$D_{n}4000$,][]{Balogh99}, which is a longer-term indicator of quiescence. \cite{Kauffmann03} found that SDSS spectroscopic data shows a bimodal distribution in $D_{n}4000$.  The first peak at $D_{n}4000 \sim 1.3$ corresponds to galaxies with mean stellar ages $\sim 1-3$ Gyr, and a second peak at $D_{n}4000 \sim 1.85$ corresponds to galaxies with mean stellar ages $\sim 10$ Gyr. We see a similar distribution in our SDSS DR16 spectroscopic catalog (Fig.~\ref{fig:sm_dn4000}), and we split the star-forming and quiescent host galaxies at $D_{n}4000 = 1.6$. This split is consistent with \cite{Kauffmann03} and has been used in other analyses of SDSS galaxies \citep[e.g.,][]{Blanton11, Tinker17}. We also investigated using a stellar mass-dependent cut between red and blue $D_{n}4000$ galaxies following \cite{Geha12}, but it did not change our results. When binning isolated hosts by either SSFR or $D_{n}4000$, we do not find any significant differences between the redshift distributions of the two bins.

\begin{figure}
    \centering
    \includegraphics[width=\columnwidth]{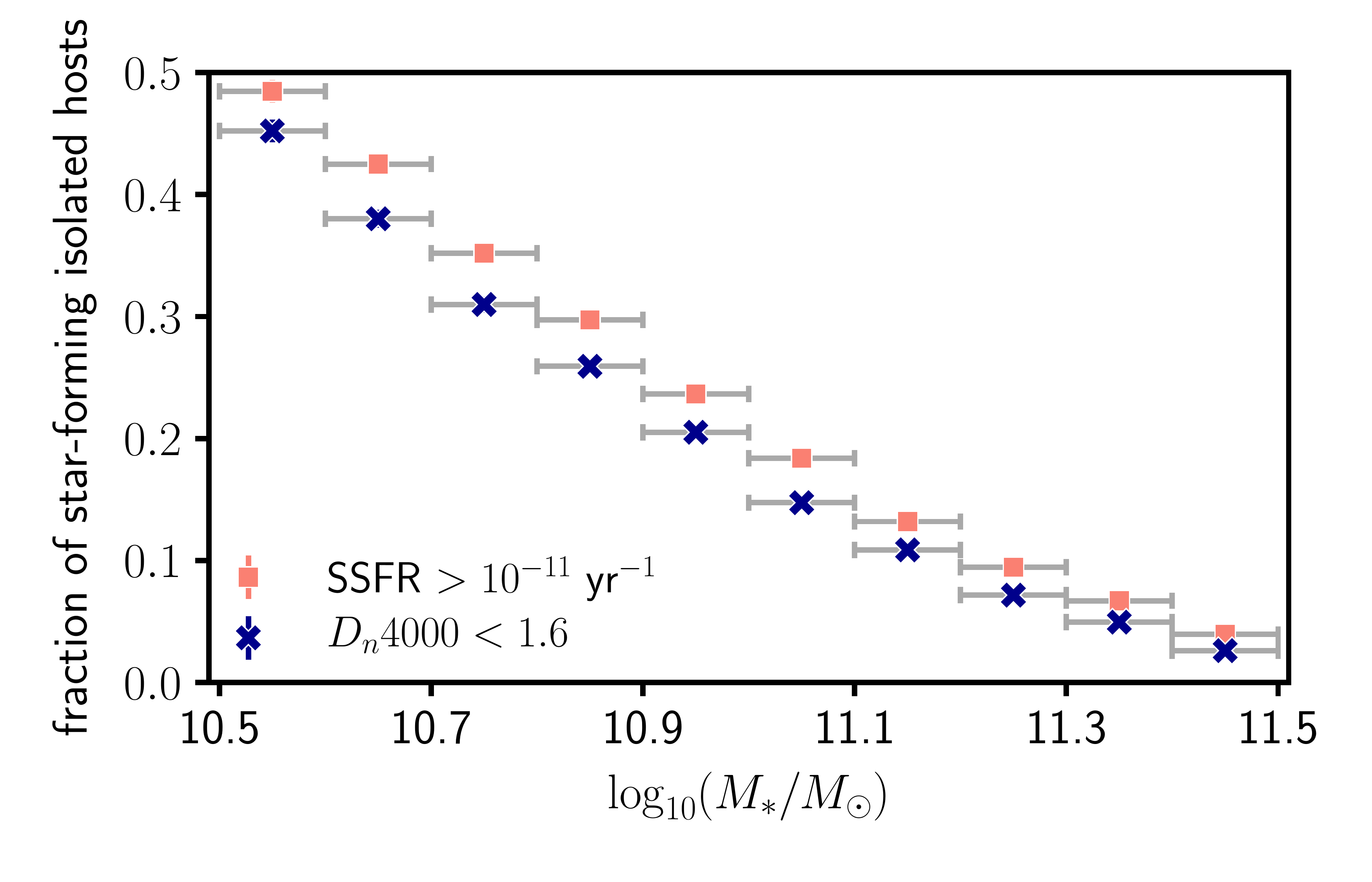}
    \caption{The fraction of star-forming isolated hosts in the SDSS is similar for both of the indicators used to bin star-forming versus quiescent hosts (specific star formation rates [SSFR] and $D_{n}4000$) across the entire isolated host mass range. Each marker indicates the star-forming fraction for isolated hosts within a 0.1 dex bin (e.g., over $10.7 < \log_{10}(M_*/M_\odot) < 10.8$). The Poisson errors in the star-forming fractions are smaller than the sizes of the plot markers, and the grey horizontal bars indicate the width of the host stellar mass bins.}
    \label{fig:sf_fraction}
\end{figure}

As in \cite{ODonnell20}, when using the star-forming fraction from SDSS to construct accretion rate correlation predictions in the simulation data from the \UM{} (\S\ref{sec:methods}), we use the SDSS star-forming fraction among isolated hosts within 0.1 dex bins (e.g., from $10.6 < \log_{10}(M_*/M_\odot) < 10.7$). Fig.~\ref{fig:sf_fraction} plots the fraction of star-forming hosts for both star formation indicators across the isolated host mass ranges. The two indicators yield similar star-forming fractions across the isolated host mass range.

\begin{figure*}
    \centering
    \includegraphics[width=\textwidth]{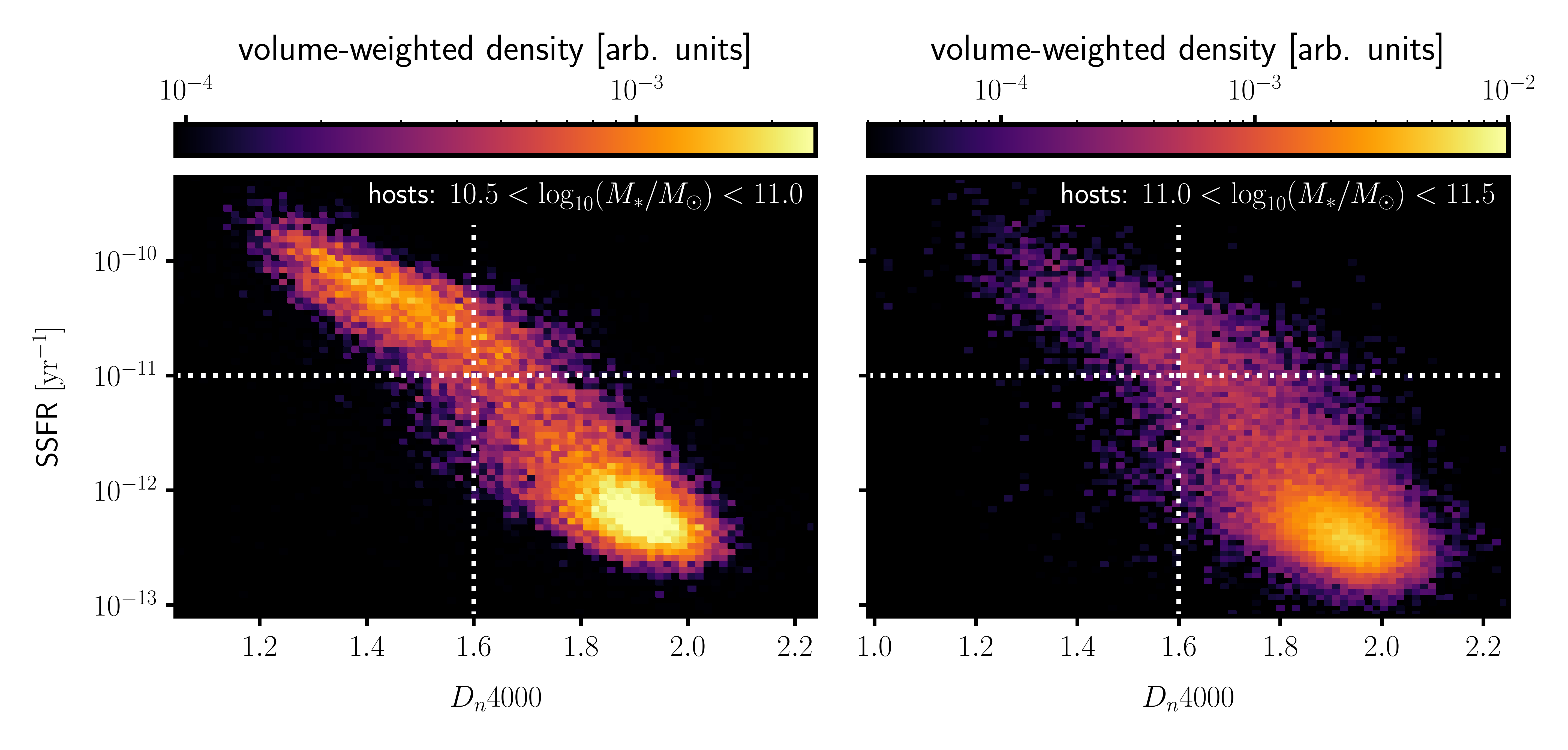}\\[-5ex]
    \caption{For isolated hosts in the SDSS, SSFR and $D_{n}4000$ measurements yield consistent bins for star-forming versus quiescent hosts. The dotted vertical and horizontal lines indicate the values used to separate isolated host mass bins for each indicator (\S\ref{sec:sf_indicators}).   The left panel shows the distribution for isolated hosts with $10.5 < \log_{10}(M_*/M_\odot) < 11.0$, and only $10.9$ per cent of hosts are classified differently between the two indicators (e.g., as star-forming by SSFR but quiescent by $D_{n}4000$). The right panel shows the distribution for isolated hosts with $11.0 < \log_{10}(M_*/M_\odot) < 11.5$, and $6.6$ per cent of hosts are classified differently between the two indicators.}
    \label{fig:SSFR_Dn4000}
\end{figure*}

Finally, Fig.~\ref{fig:SSFR_Dn4000} compares the SSFR and $D_{n}4000$ values for isolated hosts in both stellar mass bins. The two indicators track each other very well with $\lesssim 10$ per cent difference in isolated host classification. For isolated hosts with stellar masses $10.5 < \log_{10}(M_*/M_\odot) < 11.0$, 7.3 per cent of isolated hosts that are star-forming based on their SSFR values are quiescent based on the $D_{n}4000$ measurements, and 3.6 per cent of isolated hosts that are star-forming based on their $D_{n}4000$ measurements are quiescent based on their SSFR values. Similarly, for isolated hosts with stellar masses $11.0 < \log_{10}(M_*/M_\odot) < 11.5$, 4.6 per cent of isolated hosts that are star-forming based on their SSFR values are quiescent based on the $D_{n}4000$ measurements, and 2.0 per cent of isolated hosts that are star-forming based on their $D_{n}4000$ measurements are quiescent based on their SSFR values.

\subsubsection{Red vs.\ Blue Neighbours}
\label{sec:red}

As another validation of our approach, we bin neighbours by their $g-r$ colours and apply our analysis technique separately to each colour bin. As a satellite galaxy passes through the halo of its host galaxy, we expect that its star formation will quench. Galactic interactions can disturb the satellite galaxy and strip gas from the satellite halo that could otherwise replenish star formation \citep[e.g.,][]{Moore96, Gunn72, Kawata08}. Many studies have found an increase in the fraction of red galaxies within halo virial radii \citep[e.g.,][]{Dressler97, Weinmann06, Baxter17}. \cite{Wetzel13} finds that the typical timescales for quenching are on the order of satellite orbital periods (2-4 Gyr), which matches the timescales for changes in the shape of the neighbour density distributions. Thus, we expect red and blue neighbours will correspond to long and short timescales since infall, respectively.

\begin{figure}
    \centering
    \includegraphics[width=\columnwidth]{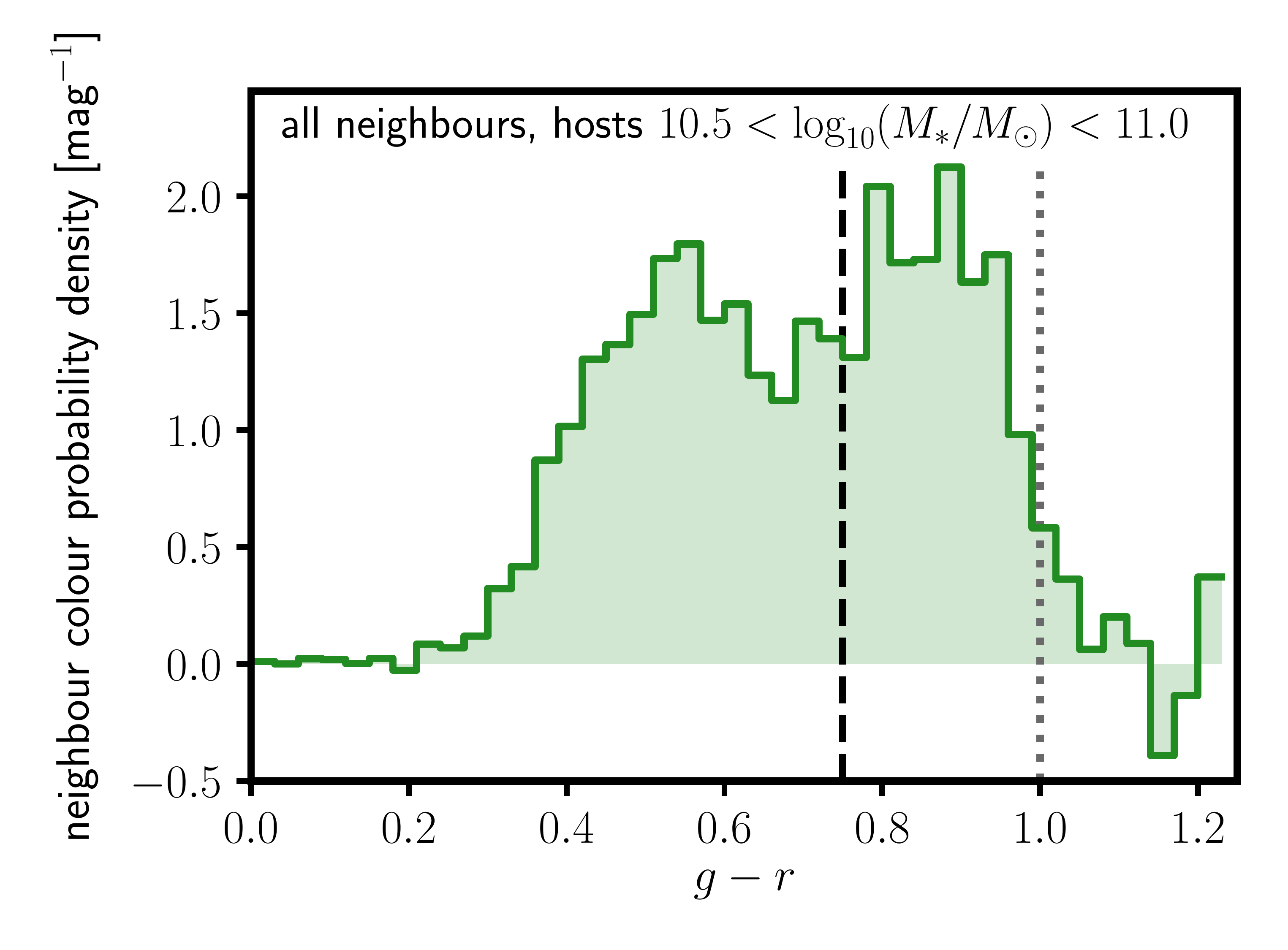}\\[-5ex]
    \caption{There are two peaks in the distribution of the $g-r$ colours of  neighbouring galaxies between 0.05-2.0 Mpc from isolated hosts with $10.5 < \log_{10}(M_*/M_\odot) < 11.0$. The dotted line indicates the $g-r < 1.0$ colour cut applied to exclude photometric galaxies from higher redshifts. The dashed line at $g-r = 0.75$ indicates the value used to separate red and blue neighbour galaxies.}
    \label{fig:neighbour_gr_redcut}
\end{figure}

We perform our analyses on both red and blue neighbours around isolated hosts with $10.5 < \log_{10}(M_*/M_\odot) < 11.0$. We define these two bins using the $g-r$ colour distribution of all neighbours within our analysis area, i.e., 0.05 - 2.0 Mpc from the isolated hosts (Fig.~\ref{fig:neighbour_gr_redcut}). We define blue neighbours as those with $0.0 < g-r < 0.75$ and red neighbours as those with $0.75 < g-r < 1.0$. 

\subsection{Simulation Data}
\label{sec:th_data}

We use haloes from the \textit{Bolshoi-Planck} dark matter simulation \citep{Klypin-BP, RodriguezPuebla-BP}, which followed a co-moving volume of (250 Mpc$/h$)$^3$ with high mass resolution ($1.6 \times 10^8 h^{-1}  \textrm{M}_\odot$) and 2048$^3$ particles ($\sim 8 \times 10^9$). \textit{Bolshoi-Planck} adopted a flat $\Lambda$CDM cosmology ($h = 0.678$, $\Omega_m = 0.307$, $\sigma_8 = 0.823$, $n_s = 0.96$); we also use this cosmology in our analysis. Halo finding and merger tree construction were done with \textsc{Rockstar} \citep{Behroozi_rockstar} and \textsc{Consistent Trees} \citep{Behroozi_consistenttrees}, respectively. Following \cite{ODonnell20}, halo accretion rates are derived from \textit{Bolshoi-Planck} over the past dynamical time $t_{\mathrm{dyn}} = 1/\sqrt{G\rho_{\mathrm{vir}}}$. We use \textit{specific} halo mass accretion rates, which are normalised by halo virial masses, i.e.,
\begin{equation}
\Gamma = \frac{\Delta \log (M_\mathrm{vir})}{\Delta \log (a)} \equiv \frac{\log \Big( \frac{M_\mathrm{vir}(t_\mathrm{now})}{M_\mathrm{vir}(t_\mathrm{now}-t_\mathrm{dyn})}\Big) }{\log \Big( \frac{a(t_\mathrm{now})}{a(t_\mathrm{now}-t_\mathrm{dyn})}\Big) }\, ,
\label{eq:accrate}
\end{equation}
following \cite{Diemer14}. The distribution of these accretion rates only weakly depends on halo mass \citep{Behroozi15b}. In \cite{ODonnell20}, we also used specific halo accretion rates calculated over the past $2t_{\mathrm{dyn}}$ and found they were consistent with results using specific halo accretion rates over $1t_\mathrm{dyn}$ \citep[Appendix B of][]{ODonnell20}.

For galaxy stellar masses, we use those from the \UM{} empirical model \citep{Behroozi19}, which implemented a Markov Chain Monte Carlo (MCMC) algorithm to model relationships between dark matter halo properties and galaxy properties \citep{Behroozi19}. The \UM{} self-consistently constrained individual galaxies' properties to match observed stellar mass functions ($z \sim 0- 4$), specific star formation rates ($z \sim 0-8$), cosmic star formation rates ($z\sim 0 -10$), UV luminosity functions ($z \sim 4 -10$), median UV-stellar mass relations ($z \sim 4-10$), auto- and cross-correlation functions ($z \sim 0-0.5$), and quenched fractions ($z\sim 0-4$). The \UM{} constrained stellar masses at $z=0$ to match \cite{Moustakas13} and used corrections from \cite{Bernardi13} for extended galaxy profiles. Additionally, the \UM{} allowed for orphans, i.e., it allowed satellites to persist after being destroyed in the dark matter simulation. Without including orphans, the model would predict a lower galaxy spatial correlation than is observed (see Appendix C of \citealt{Behroozi19} and \S{}2.2.2. of \citealt{Allen19}). In \cite{ODonnell20}, we tested our results by excluding the orphan model and found that while it did slightly affect the neighbour density distributions close to the isolated hosts, it did not significantly alter our results \citep[Appendix C of][]{ODonnell20}.

Following \cite{ODonnell20}, we use galaxy positions and velocities from the \UM{}. We also use observed stellar masses from the \UM{}, which incorporate both random scatter and systematic offsets. While the \UM{} also generates star formation rates, we discard this information to allow choosing SFRs that have different correlations with halo accretion rates.

\subsubsection{Sample Statistics}
\label{sec:th_stats}

As in \cite{ODonnell20}, we combined catalogs from 14 simulation snapshots with $a = 0.904$ to $a = 1.002$. We identified isolated hosts following the same criteria as the observational data (no galaxy with a higher observed stellar mass within 2 Mpc projected distance and 1000 km/s velocity separation). Each snapshot had an average of 31026 isolated hosts with $10.5 < \log_{10}(M_*/M_\odot) < 11.0$ and 9541 isolated hosts with $11.0 < \log_{10}(M_*/M_\odot) < 11.5$. We note that $\gtrsim 94$ per cent of the isolated hosts were not satellites of larger haloes for both isolated host mass bins.\footnote{The isolated galaxy catalogs are available at \url{https://github.com/caodonnell/DM_accretion}}

To measure the uncertainties in the neighbour density distributions, we use jackknife sampling. We created 25 jackknife samples by averaging across the 14 snapshots with the same  50$\times$50 Mpc region removed from each snapshot. Each jackknife sample has an average of $\sim$27000 isolated hosts with $10.5 < \log_{10}(M_*/M_\odot) < 11.0$ and $\sim$9000 isolated hosts with $11.0 < \log_{10}(M_*/M_\odot) < 11.5$. As noted in \cite{ODonnell20}, the uncertainties for \UM{} results differ from those for SDSS results because the background (noise) from the SDSS photometric data includes galaxies out to $z \sim 0.2$ (over 570 Mpc/$h$). However, the \UM{} simulation box is only 250 Mpc/$h$ per side.

\section{Methods}
\label{sec:methods}

Our methodology follows the technique described in \cite{ODonnell20}. Briefly, we identify isolated galaxies from  SDSS spectroscopic data \citep{SDSS_DR16} with no larger neighbouring galaxy within 2 Mpc projected (on-sky) physical distance or 1000 km/s velocity distance. We term these galaxies our \textit{isolated host} sample. We calculate the density distribution of neighbouring galaxies using SDSS photometric data \citep{SDSS_DR16}. To eliminate background and foreground contamination, for each isolated host, we create 100 random pointings that also follow our isolation criteria within the same sky footprint, and we subtract the neighbour density distribution around random pointings from the neighbour density distribution around our isolated hosts. We replicate this procedure in our simulation data from \UM{} snapshots \citep{Behroozi19} by identifying isolated haloes, calculating the density of nearby neighbours, and subtracting background and foreground contamination by using 100 random pointings per isolated host.

Additionally, our methodology accounts for systematic biases in our data \citep[\S 2.3 in][]{ODonnell20}. First, the stellar mass function from the \UM{} differs from the stellar mass function in the SDSS MPA-JHU value-added catalogue due to different assumptions in converting luminosities to stellar masses; these result in the \UM{} having more high-mass galaxies ($M_* > 10^{11} M_\odot$; Fig. 10 in \cite{ODonnell20}). We account for these differences by choosing analogous stellar mass cutoffs in the \UM{} such that the cumulative number density of galaxies with greater stellar masses matches that from the SDSS MPA-JHU catalogue (Table \ref{tab:MF}). 

Second, the \UM{} assumes that the observed stellar masses of quiescent and star-forming galaxies have the same biases, but this may not be true in the real Universe given differences in  metallicity, dust, and star formation histories between the two populations. These differences create a normalisation offset in the neighbour density distributions, though it should not affect the shapes of the distributions. We calculate this offset by matching the neighbour density distributions from SDSS and \UM{} between 1.25-2.0 Mpc, as this region has the least correlation with accretion rates \citep{ODonnell20}. Table \ref{tab:density_norm} lists typical values for these normalisation factors. In \cite{ODonnell20}, we thoroughly tested the validity of this approach by comparing our observational results to simulation data from the \UM{} where we selected simulated isolated hosts such that their density distribution normalisations matched the observed values \citep[\S{}4.2 of][]{ODonnell20}. We conducted this test using the same jackknife sampling across \UM{} catalogs as is used in this paper, as well as using a single snapshot ($a=0.956$) with haloes from the Small MultiDark Planck simulation \citep[SMDPL;][]{Klypin16}, which has a larger co-moving volume than the \textit{Bolshoi-Planck} simulation ((400 Mpc$/h$)$^3$ versus (250 Mpc$/h$)$^3$, respectively). Both tests yielded consistent results with our approach applying calculated normalisation factors between the SDSS and \UM{} neighbour density distributions. We also conducted additional tests assuming stellar mass offsets in the SDSS values (Appendix D1) and the \UM{} values (Appendix D2) to ensure the robustness of our analysis technique \citep{ODonnell20}.

Finally, to account for stellar mass completeness and background/foreground projection effects in SDSS, we weight the neighbour density distributions from SDSS data by
\begin{equation}
    w = z^2 \times \frac{1}{V_\mathrm{max}(M_*)} \, .
\end{equation}
The factor of $z^2$ maximises signal-to-noise given Poisson variance in unassociated source counts (which scales as $z^{-2}$), and the factor $1/V_\mathrm{max}(M_*)$ accounts for stellar mass completeness as computed from Eq.~\ref{eq:SM_complete_cut}. For a more detailed description of these weights, see \S 2.3.3 of \cite{ODonnell20}. We also reported in \cite{ODonnell20} that our results did not change if we weighted the distributions only by stellar mass completeness and exclude the inverse variance weights.

\begin{table}
\begin{center}
\begin{tabular}{c|c|c|c}
    &SDSS  & \UM{} & $\Phi(>M_*)$ \\
    &$\log_{10}(M_*/M_\odot)$ & $\log_{10}(M_*/M_\odot)$ & (Mpc/$h$)$^{-3}$ \\
    \hline 
    \multirow{3}{*}{\rotatebox{90}{Hosts}} 
    & 10.50 & 10.50 & 0.64016  \\
& 11.00 & 11.08 & 0.09464  \\
& 11.50 & 11.75 & 0.00207 \\
\hline \hline
\multirow{4}{*}{\rotatebox{90}{Neighbours}}
&8.50 & 8.62 & 6.62222  \\
&9.00 & 8.93 & 4.85279  \\
&9.50 & 9.38 & 3.05361  \\
&10.00 & 9.93 & 1.62929 \\
\hline
\end{tabular}
\caption{The SDSS MPA-JHU catalogue and \UM{} include different assumptions that affect their stellar mass functions.  In our analysis, we use analogous stellar mass cutoffs in the \UM{} such that the cumulative number density of more massive objects matches that from SDSS MPA-JHU. The first two columns summarises these stellar masses, and the third column indicates the cumulative number density of more massive galaxies. The first three rows are the limits for selecting isolated hosts, and the bottom four rows are the values for selecting nearby neighbours.  Throughout this paper, we use stellar masses from the SDSS (first column).}
\label{tab:MF}
\end{center}
\end{table}

\begin{table}
\begin{center}
\begin{tabular}{c|c|c}
   Isolated Host Mass & Star-Formation & Normalisation Factor \\
   
   $[\log_{10}(M_*/M_\odot)]$ & Indicator & $[$dex$]$ \\ \hline
   \multirow{4}{*}{(10.5, 11.0)} & \textcolor{blue}{SSFR $> 10^{-11}$ yr$^{-1}$}& $-0.226 \pm 0.206$ \\
& \textcolor{red}{SSFR $< 10^{-11}$ yr$^{-1}$}& $-0.020 \pm 0.088$ \\
& \textcolor{blue}{$D_{n}4000 < 1.6 $}& $-0.199 \pm 0.208$ \\
& \textcolor{red}{$D_{n}4000 > 1.6$}& $-0.035 \pm 0.086$ \\ 

\hline
   
   \multirow{4}{*}{(11.0,11.5)} & \textcolor{blue}{SSFR $> 10^{-11}$ yr$^{-1}$}& $-0.046 \pm 0.109$ \\
& \textcolor{red}{SSFR $< 10^{-11}$ yr$^{-1}$}& $0.033 \pm 0.033$ \\
& \textcolor{blue}{$D_{n}4000 < 1.6 $}& $0.094 \pm 0.090$ \\
& \textcolor{red}{$D_{n}4000 > 1.6$}& $0.017 \pm 0.033$ \\ 

\hline

\end{tabular}
\caption{Following \protect\cite{ODonnell20}, we apply a normalisation correction to match the neighbour density distributions between the \UM{} and SDSS between 1.25-2.0 Mpc. This factor is required because the \UM{} assumes the same biases between true and observed stellar masses for both star-forming and quiescent hosts. This table summarises the average normalisation factors between the observed SDSS neighbour density distributions and the \UM{} predictions for no correlation ($\rho=0$) between dark matter accretion rates and star formation rates. We include both star formation and quiescence indicators used in this paper. SSFR $>10^{-11}$ yr$^{-1}$ or $D_{n}4000 < 1.6$ selects \textcolor{blue}{star-forming hosts}, and SSFR $<10^{-11}$ yr$^{-1}$ or $D_{n}4000 > 1.6$ selects \textcolor{red}{quiescent hosts}. For isolated hosts with $10.5 < \log_{10}(M_*/M_\odot) < 11.0$, we average the results for neighbour selection limits $\log_{10}(M_*/M_\odot) > 10.0$, 9.5, and 9.0 as we are only complete down to $\log_{10}(M_*/M_\odot) > 8.95$ at the median host redshift $z=0.079$. For isolated hosts with $11.0 < \log_{10}(M_*/M_\odot) < 11.5$, we average the results for neighbour selection limits $\log_{10}(M_*/M_\odot) > 10.0$ and 9.5 as we are only complete down to $\log_{10}(M_*/M_\odot) > 9.30$ at the median host redshift $z=0.116$.}
\label{tab:density_norm}
\end{center}
\end{table}

As we demonstrated in \cite{ODonnell20}, the shapes of the neighbour density distributions encode information about $\rho$, the correlation between dark matter accretion and star formation. Specifically, the neighbour density distributions around highly-accreting hosts steepen at a few hundred kpc, consistent with expectations that newly accreted dark matter is deposited at $\sim R_{200\mathrm{m}}$ \citep{WetzelNagai15, Diemer13}. To quantify this shape and compare neighbour density distributions, we defined a \textit{shape parameter} metric \citep[\S~2.2 of][]{ODonnell20}:
\begin{equation}
    R = \frac{N \in (0.05 \textrm{ Mpc} - r_{\mathrm{split}})}{N \in (r_{\mathrm{split}} - 2.0  \textrm{ Mpc})} \, ,
    \label{eq:shape_ratio}
\end{equation}
which compares the number of neighbours close to isolated host galaxies versus the number of neighbours further away. The inner radius (0.05 Mpc) conservatively excludes incompleteness from source blending in SDSS data, and the outer radius (2.0 Mpc) is consistent with our isolation criterion. In \cite{ODonnell20}, we also repeated our analysis with an even more conservative inner radius of 0.125 Mpc and found that it did not significantly change our results.

We determined that $r_\mathrm{split} \equiv 0.316$ Mpc maximises our sensitivity to differences between host halo dark matter accretion rates for star-forming and quiescent galaxies \citep{ODonnell20}. We quantify these differences using a \textit{shape ratio} $R_\mathrm{SF}/R_\mathrm{Q}$ to compare the shape parameters of star-forming galaxies ($R_\mathrm{SF}$) versus quiescent galaxies ($R_\mathrm{Q}$). In our analysis of the SDSS data, we compute confidence levels by comparing the observed shape ratios to $R_\mathrm{SF}/R_\mathrm{Q} \equiv 1.0$, which is the expected shape ratio for no correlation ($\rho = 0$) between dark matter accretion and star formation activity. 


\begin{figure*}
    \centering
    \includegraphics[width=0.9\textwidth]{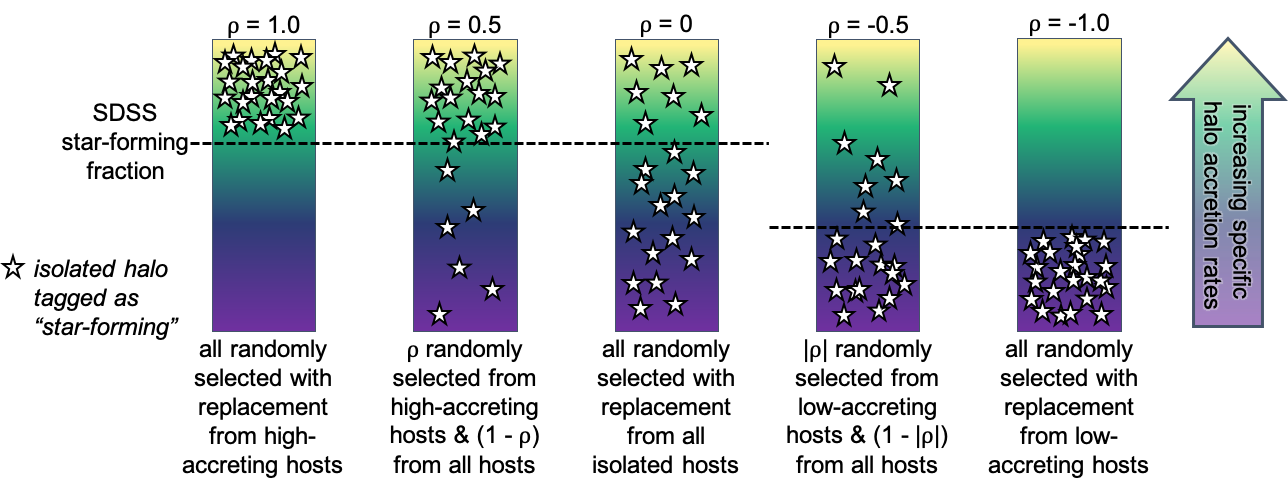}
    \caption{Schematic for generating analogues to the star-forming SDSS isolated hosts from \UM{} data to test different correlation strengths $\rho$ between dark matter accretion and star formation. The coloured bars indicate all of the isolated hosts identified in the \UM{} within a 0.1 dex stellar mass bin (e.g., $10.7 < \log_{10}(M_*/M_\odot) < 10.8$) sorted by increasing specific halo accretion rate (Eq.~\ref{eq:accrate}) from the bottom (purple) to the top (yellow) of the bars. The star icons in each example depict an isolated halo that is tagged as `star-forming.' The dashed horizontal line indicates the corresponding star-formation fraction from the SDSS within the isolated host mass bin. For positive correlations ($\rho > 0$), this fraction is applied to identify the highest-accreting hosts; for negative correlations ($\rho < 0$), this fraction is applied to identify the lowest-accreting hosts. A similar strategy is used to create analogues to the quiescent isolated hosts from the SDSS.}
    \label{fig:acc_schematic}
\end{figure*}

To construct our dark matter accretion predictions, we bin isolated hosts from the \UM{} simulation data (\S~\ref{sec:th_data}) by their specific halo accretion rates (Eq.~\ref{eq:accrate}) to match the fraction of star-forming versus quiescent hosts in SDSS. This procedure is described in detail in \S{}3.3.3 of \cite{ODonnell20} and is summarised here in Fig.~\ref{fig:acc_schematic}. Additionally, in \cite{ODonnell20}, we showed that this procedure recovers the expected correlation strength $\rho$ between specific dark matter accretion rates and a host's status as star-forming or quiescent \citep[Fig. 6 in ][]{ODonnell20}. Briefly, we split the isolated host sample into 0.1 dex-wide bins of stellar mass to calculate the star-forming fraction, e.g., in the SDSS, for isolated hosts with $10.7 < \log_{10}(M_*/M_\odot) < 10.8$, 31 per cent are star-forming based on having $D_{n}4000 < 1.6$. We split the isolated hosts from the \UM{} data into high- and low-accreting subsamples (based on their specific halo accretion rates) such that they match the star-forming fraction from SDSS for the relevant 0.1-dex bin of isolated host stellar mass. We then create `star-forming' and `quiescent' analogues using the correlation strength $\rho$ between halo accretion rates and star formation rates. For positive correlations, the star-forming analogues have a fraction $\rho$ of hosts randomly selected with replacement from the high-accreting subsample, and quiescent analogues have a fraction $\rho$ randomly selected with replacement from the low-accreting subsample. For negative correlations, the star-forming analogues have a fraction $|\rho|$ randomly selected with replacement from the low-accreting subsample, and the quiescent analogues have a fraction $|\rho|$ randomly selected with replacement from the high-accreting subsample. The remaining fraction of hosts in the analogues ($1-|\rho|$) are randomly selected with replacement from all isolated hosts identified in the \UM{}.

\section{Results}
\label{sec:results}

Below, we present results to test our choice of star formation indicator (\S\ref{sec:results_sf}), results from splitting neighbouring galaxies into red and blue subsamples (\S\ref{sec:results_red}), and results from a higher isolated host mass range (\S\ref{sec:results_hostmass}). In Appendix \ref{sec:appendix_neighbourM}, we include plots for other neighbour selection limits.

\subsection{Star Formation \& Quiescence Indicators}
\label{sec:results_sf}

\begin{figure*}
    \centering
    \includegraphics[width=\textwidth]{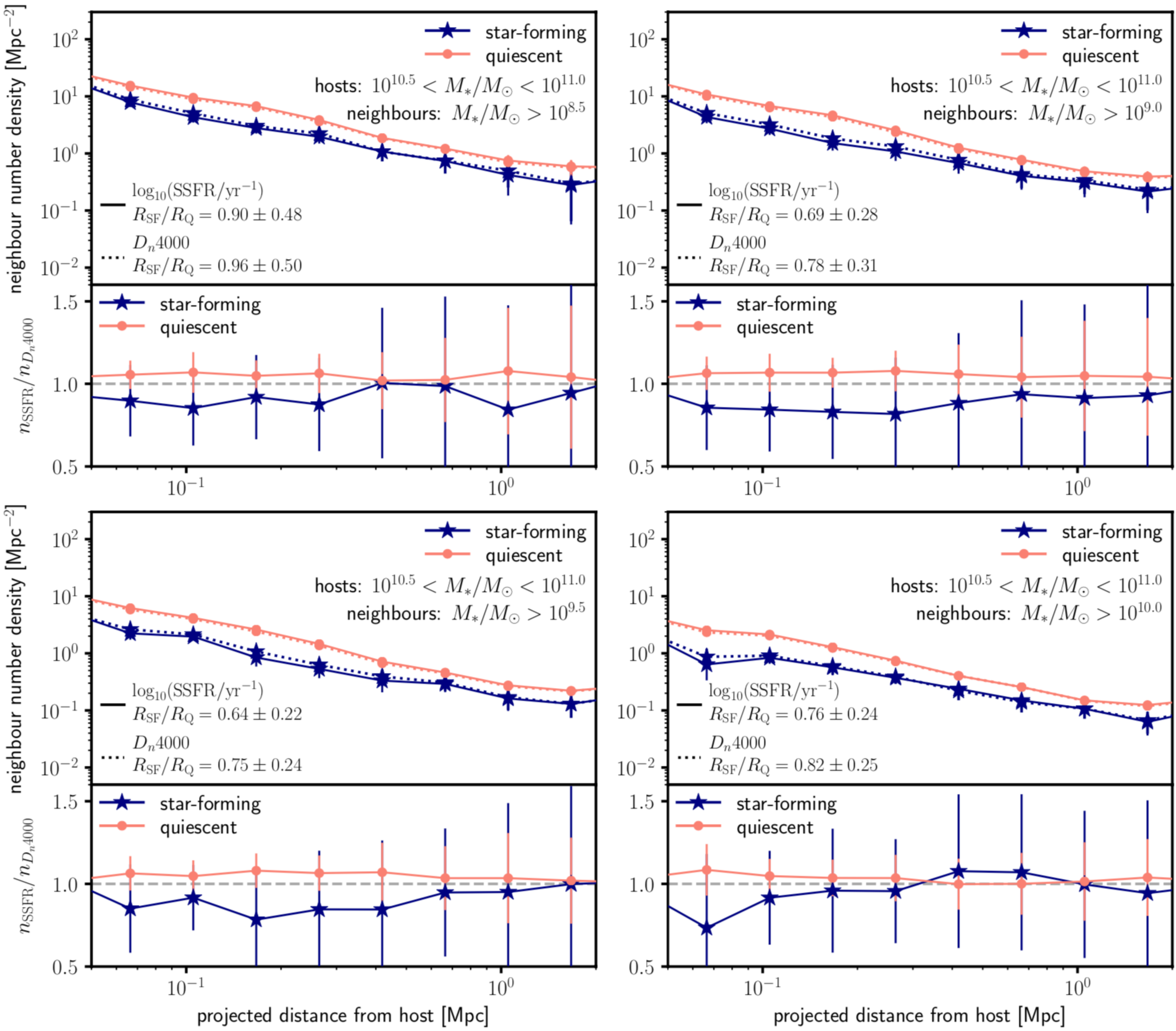}
    \caption{The neighbour density distributions around isolated hosts with $10.5 < \log_{10}(M_*/M_\odot) < 11.0$ are very similar when binning hosts by SSFRs or $D_{n}4000$, due to the fact that the two indicators are highly correlated among our isolated host sample (Fig.\ \ref{fig:SSFR_Dn4000}). The four panels represent different neighbour mass selection limits. In each panel, the top plots compares the neighbour density distributions when binning isolated hosts by SSFR versus $D_{n}4000$. The bottom plots shows the ratio of those distributions with a dashed horizontal line at $n_\mathrm{SSFR}/n_{D_{n}4000} = 1$ as a visual guide.}
    \label{fig:results_sf}
\end{figure*}

Fig.~\ref{fig:results_sf} compares the neighbour density distributions around isolated hosts with $10.5 < \log_{10}(M_*/M_\odot) < 11.0$ when binned by SSFRs versus $D_{n}4000$ for the different neighbour mass selection limits. Since $\gtrsim 90$ per cent of isolated hosts were binned in the same way by the two indicators (Fig.~\ref{fig:SSFR_Dn4000}), the resulting neighbour density distributions are also very similar. For both indicators, we see a dip in the neighbour density distribution for neighbours with higher masses ($M_* \gtrsim 10^{9.5} M_\odot$) at $< 0.1$ Mpc from the isolated hosts. 
We find consistent neighbour density distributions and shape ratios ($\Delta R_\mathrm{SF}/R_\mathrm{Q} \sim 0.2\sigma$) when binning the isolated hosts based on either SSFR or $D_{n}4000$. Both indicators yield results that are consistent with non-positive correlations ($\rho \leq 0$) between dark matter accretion and star formation at $\gtrsim 75$ per cent confidence.

\subsection{Neighbour Colours}
\label{sec:results_red}

Table \ref{tab:red_fraction} reports the fraction of red neighbours (i.e., $0.75 < g-r < 1.0$) around isolated hosts with $10.5 < \log_{10}(M_*/M_\odot) < 11.0$. There is a small, though statistically insignificant, decrease in the fraction of red neighbours with $M_* > 10^{10} M_\odot$ as the distance from the isolated host increases. The difference in the fraction of red neighbours around star-forming versus quiescent isolated hosts is not statistically different due to large uncertainties (\S\ref{sec:obs_stats}).

\begin{figure*}
    \centering
    \includegraphics[width=\textwidth]{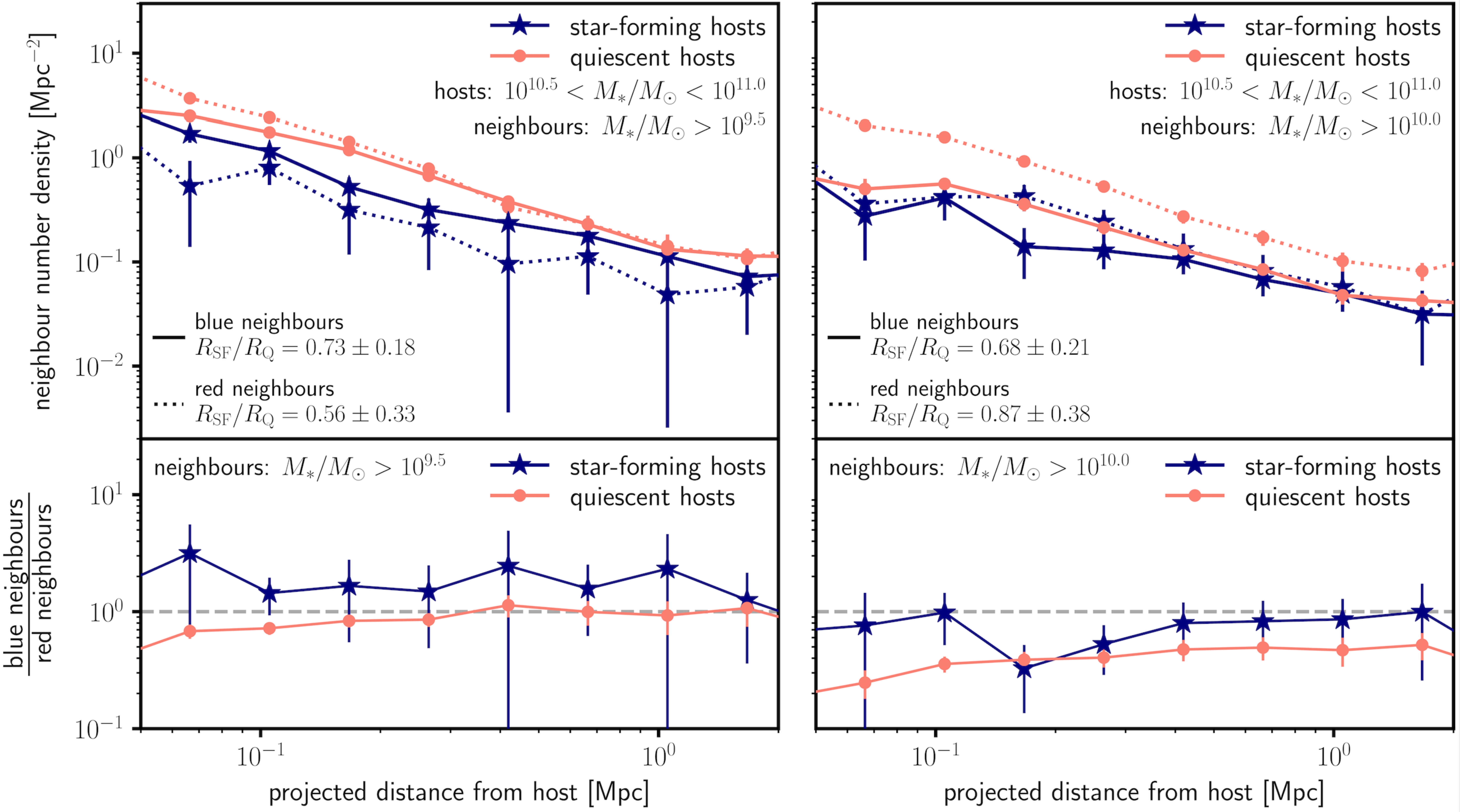}
    \caption{The neighbour density distributions around red and blue isolated hosts are very similar, suggesting that our finding of correlation strengths $\rho \leq 0$ (based on the measured shape ratios being $R_\mathrm{SF}/R_\mathrm{Q} < 1$; see Fig.~\ref{fig:results_rsf_rq_neighbours} below) between dark matter accretion and star formation applies to both recent and older infall populations as traced by blue and red neighbours, respectively.  The star-forming and quiescent hosts are separated based on their SSFRs (\S\ref{sec:sf_indicators}). The top panels compare the neighbour density distributions of red and blue neighbours around isolated hosts with $10.5 < \log_{10}(M_*/M_\odot) < 11.0$, and the bottom row indicates the ratio of the blue neighbour density distribution to the red neighbour density distribution. In the bottom row, a horizontal dashed line at $n_\mathrm{blue}/n_\mathrm{red} = 1.0$ is included as a visual guide for the slope of the ratio as a function of projected distance, although we do not necessarily expect the value of the observed ratio to equal 1.}
    \label{fig:results_red}
\end{figure*}

\begin{figure}
    \centering
    \includegraphics[width=\columnwidth]{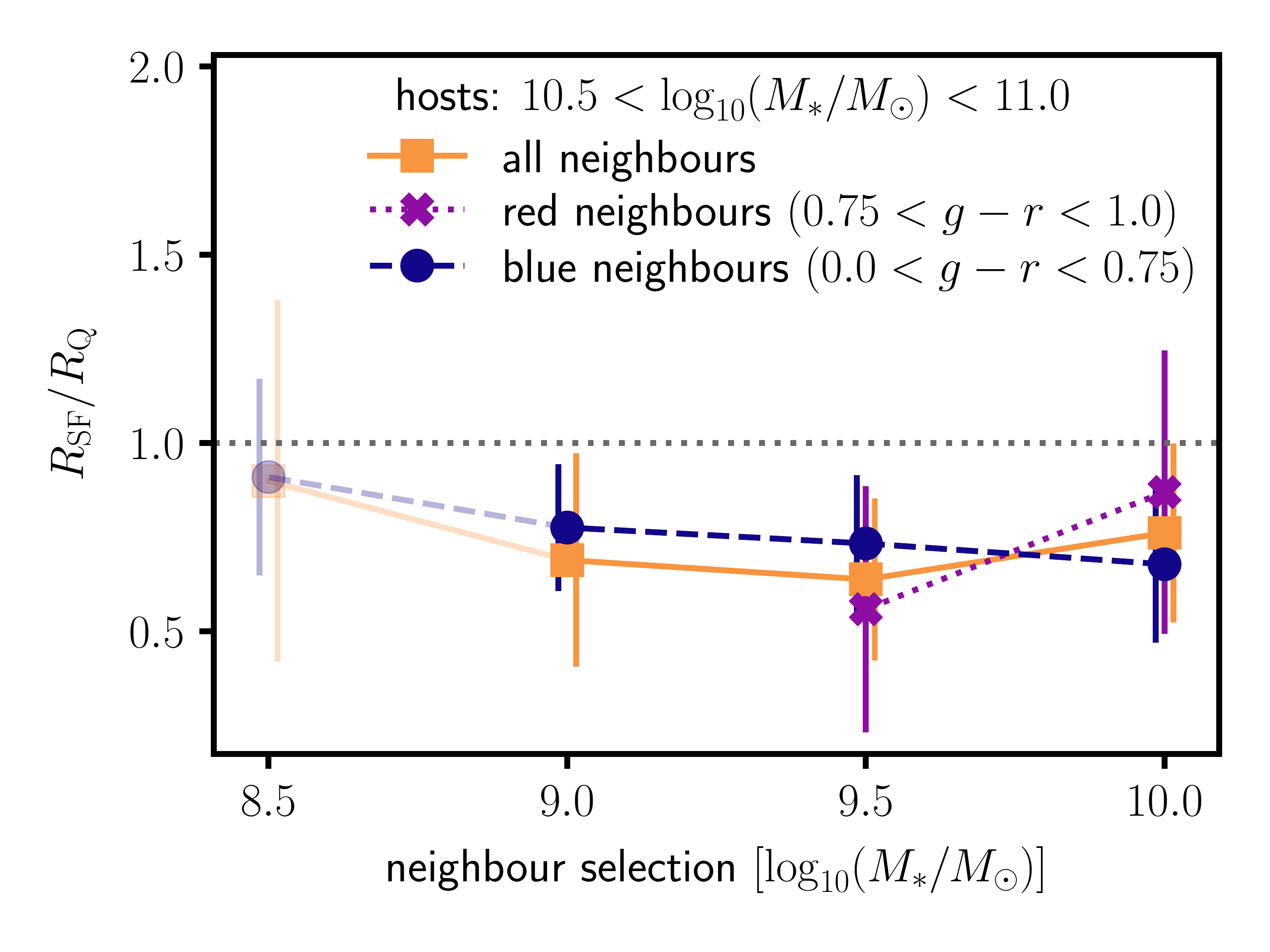}\\[-5ex]
    \caption{The shape ratio is not consistent with positive correlations between halo accretion rates and star formation regardless of nearby neighbour colours. Since we expect red neighbours may probe longer timescales than blue neighbours, these results likely mean that different infall populations are not affected in a significantly different manner by recent accretion for host galaxies in this mass range. The star-forming and quiescent hosts are separated based on their SSFRs (\S\ref{sec:sf_indicators}). We only plot shape ratios for red neighbours ($0.75 < g-r < 1.0$) for neighbours with $M_* > 10^{9.5} M_\odot$ because the measurements for lower-mass neighbours are noise-dominated. The plot markers for the neighbour selection $M_* > 10^{8.5} M_\odot$ are faded because neighbours of this stellar mass are not observable for all isolated hosts; for isolated hosts with $10.5 < \log_{10}(M_*/M_\odot) < 11.0$, at the median redshift $z=0.079$, the SDSS observation limit for neighbours is $M_* > 10^{8.95} M_\odot$, and at the maximum redshift $z=0.123$, the SDSS observation limit for neighbours is $M_* > 10^{9.36} M_\odot$.}
    \label{fig:results_rsf_rq_neighbours}
\end{figure}

Fig.~\ref{fig:results_red} shows that the neighbour density distributions of red and blue neighbours ($0.75< g-r < 1.0$ and $ 0.0 < g-r < 0.75$, respectively) have similar shapes. For this analysis, we separate the isolated hosts into star-forming and quiescent bins based on their SSFRs; however, since SSFRs and $D_{n}4000$ measurements track each other very closely (Fig.~\ref{fig:SSFR_Dn4000}), the choice of star formation indicator does not yield significantly different results. Because of the minimum colour cutoff for red neighbours ($g-r > 0.75$), we limit our analysis to neighbours with $M_* > 10^{9.5} M_\odot$ and $M_* > 10^{10} M_\odot$ to have sufficient signal-to-noise. The shape ratios (Fig.~\ref{fig:results_rsf_rq_neighbours}) are also similar ($\Delta R_\mathrm{SF}/R_\mathrm{Q} \sim 0.2\sigma$) and are consistent with non-positive correlations ($\rho \leq 0$) between dark matter accretion and star formation at $\gtrsim 90$ per cent confidence. Since we expect that red neighbours correspond to an older infall population \citep[][see also the discussion above in \S 2.1.3]{Wetzel13}, these results would imply that the shape of the distribution is independent of the time since infall. 

Furthermore, in \cite{ODonnell20}, we noted that the neighbour density distributions with higher-mass neighbour selection limits had a deficit of neighbours close to the isolated hosts ($\lesssim 1.25$ kpc). In Fig.~\ref{fig:results_red}, the blue neighbour density distribution for the $M_* > 10^{10} M_\odot$ selection shows this same deficit, as do the red neighbour density distributions around star-forming hosts for both the $M_* > 10^{9.5} M_\odot$  and $M_* > 10^{10} M_\odot$ selections. This deficit may be due to satellite galaxies being disrupted by their host galaxies more quickly than predicted in the \UM{} \citep[e.g.,][]{Garrison-Kimmel17}.

\begin{table}
    \centering
    \begin{tabular}{c|c|c|c}
         Radial Range
         & \multicolumn{2}{c}{\textbf{Fraction of Red Neighbours}} & \multirow{2}{*}{$\frac{f_\mathrm{red}(\mathrm{Q})}{f_\mathrm{red}(\mathrm{SF})}$} \\
         $[$Mpc$]$ & Star-Forming Hosts & Quiescent Hosts&\\ \hline \hline
         \multicolumn{4}{c}{Neighbours with $M_* > 10^{10}M_\odot$} \\ \hline
         0.05 - 0.316 & 0.62 $\pm$ 0.08 &  0.70 $\pm$ 0.02 & 1.12 $\pm$ 0.15\\
         0.316 - 1.00 &  0.515 $\pm$ 0.10 & 0.66 $\pm$ 0.03 & 1.29 $\pm$ 0.25\\
         0.316 - 2.00 & 0.50 $\pm$ 0.10 &  0.67 $\pm$ 0.03 & 1.33 $\pm$ 0.28\\ \hline\hline
         \multicolumn{4}{c}{Neighbours with $M_* > 10^{9.5}M_\odot$} \\\hline
         0.05 - 0.316 & 0.32 $\pm$ 0.11 & 0.52 $\pm$ 0.02 & 1.63 $\pm$ 0.58\\
         0.316 - 1.00 &  0.26 $\pm$ 0.16 & 0.50 $\pm$ 0.04 & 1.93 $\pm$ 1.23\\
         0.316 - 2.00 & 0.32  $\pm$ 0.17 & 0.50 $\pm$ 0.04 & 1.57 $\pm$ 0.85\\ \hline

    \end{tabular}
    \caption{Our analysis does not have enough power to constrain differences in the fraction of red neighbours (i.e., $0.75 < g-r < 1.0$) around star-forming versus quiescent isolated hosts with stellar masses $10.5 < \log_{10}(M_*/M_\odot) < 11.0$.  The star-forming and quiescent hosts are separated based on their SSFRs (\S\ref{sec:sf_indicators}). The uncertainties are from our jackknife sampling of the SDSS data (\S\ref{sec:obs_stats}). We include data for neighbour mass selection limits of $\log_{10}(M_*) > 9.5$ and $10.0$; the lower-mass neighbour bins are noise-dominated due to few red neighbours passing these selection limits.}
    \label{tab:red_fraction}
\end{table}


\subsection{Host Stellar Masses}
\label{sec:results_hostmass}

\begin{figure*}
    \centering
    \includegraphics[width=\textwidth]{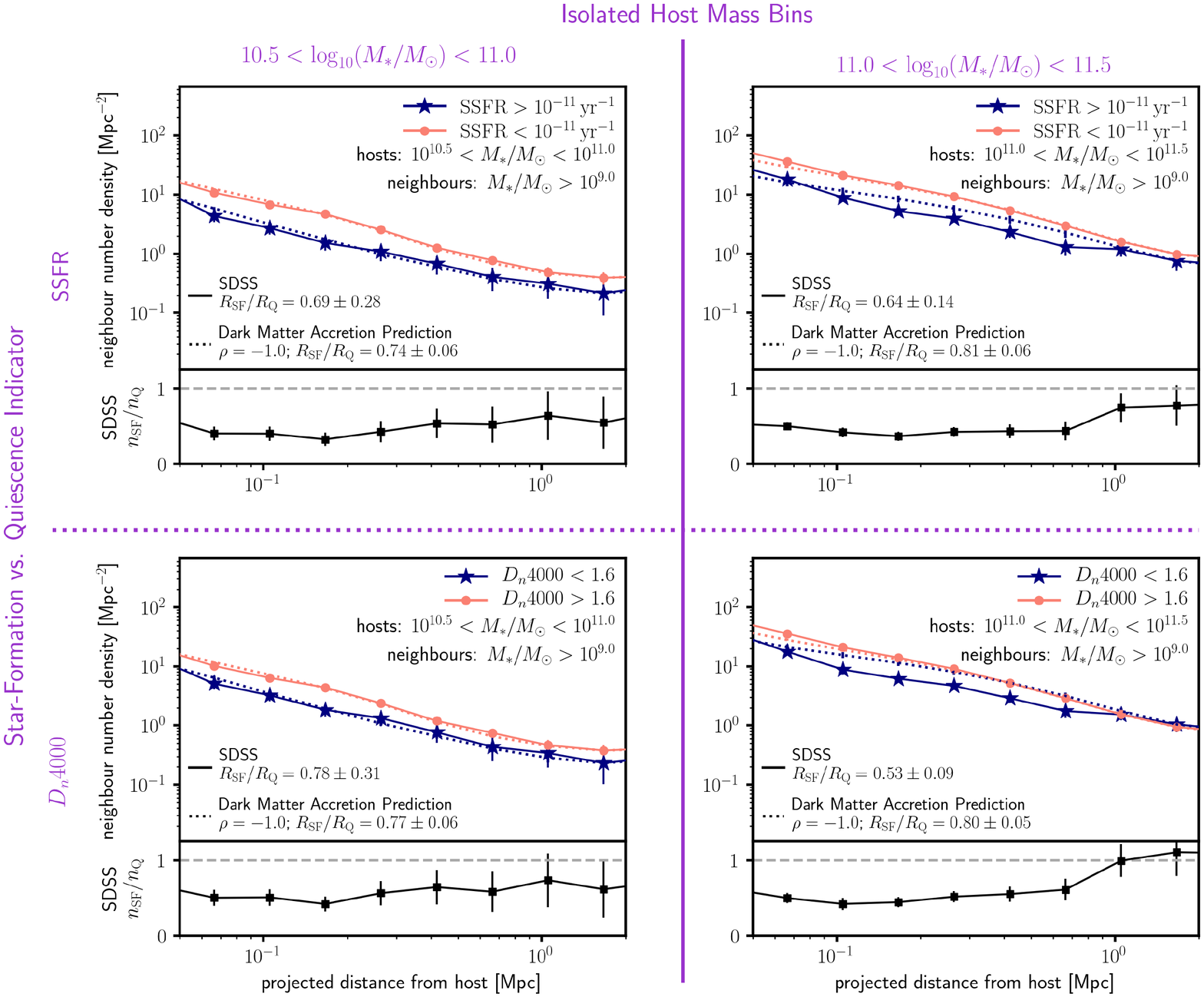}
    \caption{Our results are consistent with correlation strengths $\rho \leq 0$ between dark matter accretion and star formation regardless of isolated host mass bin (figure columns) or indicator to separate star-forming versus quiescent hosts (figure rows). This plot shows results with a neighbour mass selection $M_* > 10^{9.0} M_\odot$; Appendix \ref{sec:appendix_neighbourM} includes plots for other neighbour $M_*$ selection limits. In each panel, the top plots compare the neighbour density distributions from the SDSS to the \UM{} predictions for anti-correlation ($\rho = -1$) which is the closest match to the observed shape ratios. The bottom plots show the ratio of the observed neighbour density distributions for star-forming versus quiescent isolated hosts. A dashed horizontal line at $n_\mathrm{SF}/n_\mathrm{Q}=1$ is included as a visual guide to emphasise that the neighbour density distributions observed around star-forming hosts are flatter than the neighbour density distributions around quiescent hosts, which is consistent with non-positive correlations between dark matter accretion and star formation. }
    \label{fig:results_9.0}
\end{figure*}

\begin{figure}
    \centering
    \includegraphics[width=\columnwidth]{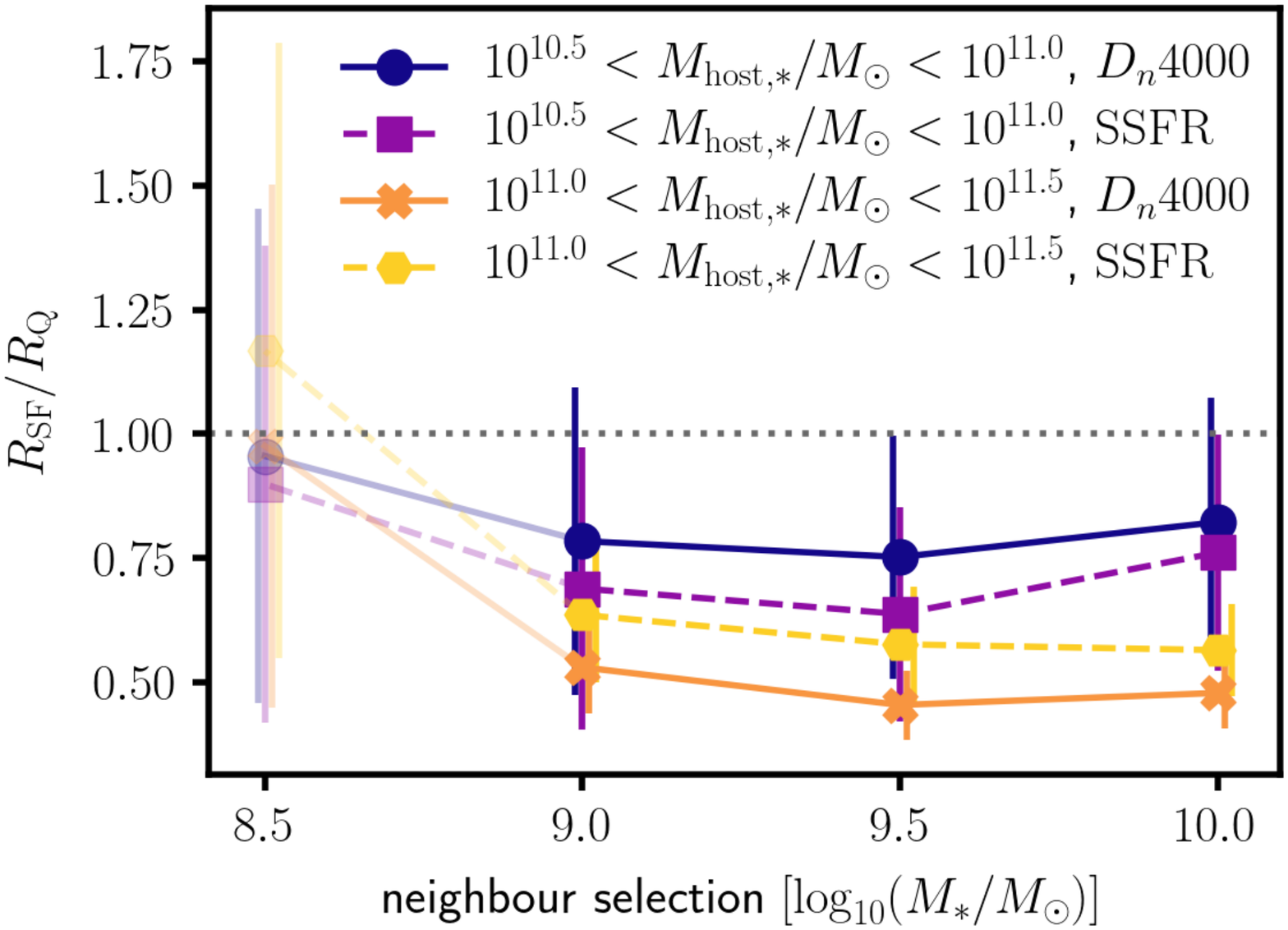} \\[-5ex]
    \caption{The shape ratios for our observed neighbour density distributions are all consistent with $\rho \leq 0$ regardless of choice of isolated host mass bin, indicator to separate star-forming versus quiescent hosts, and neighbour $M_*$ selection. As in Fig.~\ref{fig:results_rsf_rq_neighbours}, because we cannot observe neighbours with $M_* \sim 10^{8.5} M_\odot$ at all isolated host redshifts, those plot markers are shown with faded colours.}
    \label{fig:results_rsf_rq_hostbins}
\end{figure}

Finally, we compare results with higher-mass isolated hosts ($11.0 < \log_{10}(M_*/M_\odot) < 11.5$) using both star formation indicators. The rows of Fig.\ref{fig:results_9.0} (as well as Fig.~\ref{fig:results_8.5}, \ref{fig:results_9.5}, and \ref{fig:results_10.0} in Appendix \ref{sec:appendix_neighbourM}) compare across bins of isolated host mass.  Fig.~\ref{fig:results_rsf_rq_hostbins} summarises the results for both star formation indicators in the higher-mass isolated host bin. For both of our star formation indicators (SSFR and $D_{n}4000$), results are similar ($R_\mathrm{SF}/R_\mathrm{Q}$ are within $\sim 0.2-0.3\sigma$) and remain consistent with correlations $\rho \leq 0.0$ between star formation and dark matter accretion at $\gtrsim 85$ per cent confidence. We also observe a deficit in higher-mass neighbours around isolated hosts from both stellar mass bins as compared to \UM{} predictions, which may suggest that satellite galaxies are depleted by their host galaxies faster than expected \citep[e.g.,][]{Garrison-Kimmel17}.

\section{Discussion \& Conclusion}
\label{sec:disc_conclu}

We build on our work from \cite{ODonnell20}, which presented a method to constrain the correlation strength between dark matter accretion and recent star formation (as determined by SSFRs) for Milky Way-mass galaxies at $z < 0.123$ using the distribution of nearby neighbours. We found that our results favored non-positive correlations ($\gtrsim 85$ per cent confidence). In this paper, we extend this analysis by 
\begin{enumerate}
\item comparing the density distributions of red versus blue neighbours, which trace older and more recent infall populations,
\item comparing the correlation between dark matter accretion and star formation when binning isolated hosts by $D_{n}4000$ measurements, a longer-term quiescence indicator, versus binning isolated hosts by their specific star formation rates, and
\item analyzing higher-mass isolated hosts ($11.0 < \log_{10}(M_*/M_\odot) < 11.5$) as an independent check of our results.
\end{enumerate}
In all three cases, our results are consistent with non-positive correlations between dark matter accretion and star formation rates.

First, in \cite{ODonnell20}, we noted that we would expect to find weak correlations if SSFRs change on timescales much shorter than satellite orbits ($\sim 2 t_\mathrm{dyn} \sim 4$ Gyr). In this paper, we address this possible interpretation by (1) correlating dark matter accretion with $D_{n}4000$, a long-term quiescence indicator; and (2) comparing red and blue populations of nearby neighbours, which trace satellite galaxy populations with different infall timescales. All of our results are consistent with our findings in \cite{ODonnell20} that generally rule out positive correlations between dark matter accretion and star formation within SDSS observational limits.

A second consideration is that neighbouring galaxies may be a biased tracer of the host galaxies' dark matter haloes. This concern remains in this paper's analysis; for example, this bias would affect all neighbours regardless of their $g-r$ colours. Additional measurements, such as weak lensing data, are needed to provide a different tracer of dark matter haloes to test the effect of systematic biases for using neighbouring galaxies to trace the density profile.

Our results are consistent with models that invoke modest recycling timescales for ejected gas, allowing for gas to quickly cool and re-accrete onto galaxies \citep[e.g.,][]{Keres05,Dekel06, Muratov15,vandeVoort16, Nelson13, Nelson15}. These models allow for new star formation at low redshifts even in the absence of new accretion. For haloes in our isolated host sample, only $\sim 20-30$ per cent of gas is converted into stars \citep{Behroozi19}, which suggests there should be a large gas reservoir that could support further star formation. We also note the caveat that our findings would not necessarily require recycling processes if only a small fraction of accreted baryons need to be converted into star formation, thus decoupling the gas accretion from the fraction of gas that eventually gets converted into stars.

Additionally, our results remain consistent with observational studies that do not find strong positive correlations between halo growth and galaxy star formation in the recent Universe. For example, \cite{Tinker17} found only a small correlation between the fraction of quenched central galaxies in galaxy groups and their local environmental density. As well, \cite{Behroozi15} did not find a strong correlation between close galaxy pairs (a probe of major halo mergers) and star formation rates.

Our analysis of the neighbour populations did not have enough power to constrain the difference in the fraction of red versus blue neighbours around star-forming versus quiescent hosts (Table \ref{tab:red_fraction}). Previous studies have found correlations between galaxy star formation rates, colours, and morphologies between satellites and host galaxies (`one-halo' conformity, e.g., \citealt{Weinmann06}) as well as between galaxies separated at distances well beyond their virial radius (`two-halo' conformity, e.g., \citealt{Kauffmann13,Berti17}). However, \cite{Tinker18} found that measurements of two-halo conformity may be due to satellite contamination.  As our isolation criteria are extremely strict, the isolated host galaxies in our sample reside in different environments than most galaxies, and may show different conformity effects as a result \citep{Hearin16}.

Future surveys with deeper photometric or spectroscopic limits may provide a better dataset for comparing density distributions of different neighbour populations to assess two-halo conformity among isolated central galaxies at $z=0$. If evidence for two-halo conformity existed at large distances (i.e., at distances well beyond $R_\mathrm{vir}$), our finding of non-positive correlations between dark matter halo accretion rates and star formation rates would have implications for the physical origin of galactic conformity. \cite{Hearin16} found that galactic conformity could be driven by similar dark matter halo accretion rates between galaxies in the same large-scale tidal environment, but this result relied on an assumption of a strong correlation between halo accretion rates and galactic star formation. If two-halo conformity is present among isolated central galaxies, that could suggest that a different process generates these correlations between galaxy colours, star formation rates, and other properties \citep[e.g.,][]{Kauffmann15}.

Finally, as noted in \cite{ODonnell20}, future observational surveys, such as the Dark Energy Spectroscopic Instrument (DESI) Survey \citep{DESI_1science}, will allow for stronger constraints on the correlation between dark matter accretion and star formation. These surveys will detect a larger sample of isolated Milky Way-mass galaxies at higher redshifts, and thus generate a larger sample for this analysis. These data will also allow for measuring correlations between dark matter accretion and other host galaxy properties, such as metallicity, AGN activity, and velocity dispersion. 

Furthermore, these surveys will have deeper photometric and spectroscopic limits, which will improve the analyses presented in this paper. For example, we will be able to perform this analysis on lower-mass isolated hosts. With the SDSS, we could only identify a small sample of isolated hosts with $10.0 < \log_{10}(M_*/M_\odot) < 10.5$, which were noise-dominated in their neighbour density distributions (\S\ref{sec:results_hostmass}). Additionally, we were limited to binning our isolated hosts into two bins ('star-forming' and 'quiescent'). It is possible that we may see different correlation strengths between dark matter accretion and star formation among galaxies with the strongest star formation than galaxies with weaker star formation. However, we are unable to complete this analysis with enough statistical significance to compare the populations. Surveys such as DESI will allow us to identify a larger sample of these hosts and therefore have a stronger signal to measure the shapes of their neighbour density distributions. In addition, deeper photometric limits will also allow us to detect more nearby neighbours. A larger sample of these galaxies will improve the signal-to-noise level when binning nearby neighbours by colour (\S\ref{sec:results_red}) or other properties from SED fitting. These future results will provide stronger constraints on the relation between halo accretion and star formation within isolated host galaxies.

\section*{Acknowledgements}

We thank Amanda Bauer, Gurtina Besla, Marla Geha, Elisabeth Krause, Dan Marrone, and Eduardo Rozo for helpful comments during the preparation of this paper.

Support for this research came partially via program number HST-AR-15631.001-A, provided through a grant from the Space Telescope Science Institute under NASA contract NAS5-26555. PB was partially funded by a Packard Fellowship, Grant \#2019-69646.  An allocation of computer time from the UA Research Computing High Performance Computing (HPC) at the University of Arizona is gratefully acknowledged. 

The coding and plots created for this work were done with Python packages NumPy \citep{numpy, numpy_array} and Matplotlib \citep{matplotlib}

The Bolshoi simulations have been performed within the Bolshoi project of the University of California High-Performance AstroComputing Center (UC-HiPACC) and were run at the NASA Ames Research Center. Funding for the Sloan Digital Sky Survey IV has been provided by the Alfred P. Sloan Foundation, the U.S. Department of Energy Office of Science, and the Participating Institutions. SDSS-IV acknowledges
support and resources from the Center for High-Performance Computing at
the University of Utah. The SDSS web site is www.sdss.org.
SDSS-IV is managed by the Astrophysical Research Consortium for the 
Participating Institutions of the SDSS Collaboration including the 
Brazilian Participation Group, the Carnegie Institution for Science, 
Carnegie Mellon University, the Chilean Participation Group, the French Participation Group, Harvard-Smithsonian Center for Astrophysics, 
Instituto de Astrof\'isica de Canarias, The Johns Hopkins University, Kavli Institute for the Physics and Mathematics of the Universe (IPMU) / 
University of Tokyo, the Korean Participation Group, Lawrence Berkeley National Laboratory, 
Leibniz Institut f\"ur Astrophysik Potsdam (AIP), 
Max-Planck-Institut f\"ur Astronomie (MPIA Heidelberg), 
Max-Planck-Institut f\"ur Astrophysik (MPA Garching), 
Max-Planck-Institut f\"ur Extraterrestrische Physik (MPE), 
National Astronomical Observatories of China, New Mexico State University, 
New York University, University of Notre Dame, 
Observat\'ario Nacional / MCTI, The Ohio State University, 
Pennsylvania State University, Shanghai Astronomical Observatory, 
United Kingdom Participation Group,
Universidad Nacional Aut\'onoma de M\'exico, University of Arizona, 
University of Colorado Boulder, University of Oxford, University of Portsmouth, 
University of Utah, University of Virginia, University of Washington, University of Wisconsin, 
Vanderbilt University, and Yale University.

\section*{Data Availability}
No new data were generated or analysed in support of this research. The isolated host catalogs derived from SDSS DR16 \citep{SDSS_DR16} and from the \UM{} \citep{Behroozi19} are available at \url{https://github.com/caodonnell/DM_accretion}.



\bibliographystyle{mnras}
\bibliography{paper}


\bsp	

\appendix
\section{Density Distributions for Additional Neighbour Selection Limits}
\label{sec:appendix_neighbourM}

Below, we present the neighbour density distributions around isolated hosts from both mass bins ($10.5 < \log_{10}(M_*/M_\odot) < 11.0$ and $11.0 < \log_{10}(M_*/M_\odot) < 11.5$) and both indicators used to separate star-forming and quiescent hosts (SSFR and $D_{n}4000$) for the neighbour mass selections not included in Fig.\ref{fig:results_9.0} ($M_* > 10^{8.5} M_\odot$, $> 10^{9.5} M_\odot$, and $> 10^{10.0} M_\odot$). These plots follow the same plot styles as Fig.~\ref{fig:results_9.0}, and all are consistent with correlation strengths $\rho \leq 0$ between dark matter accretion and star formation (Fig.~\ref{fig:results_rsf_rq_hostbins}).

\begin{figure*}
    \centering
    \includegraphics[width=\textwidth]{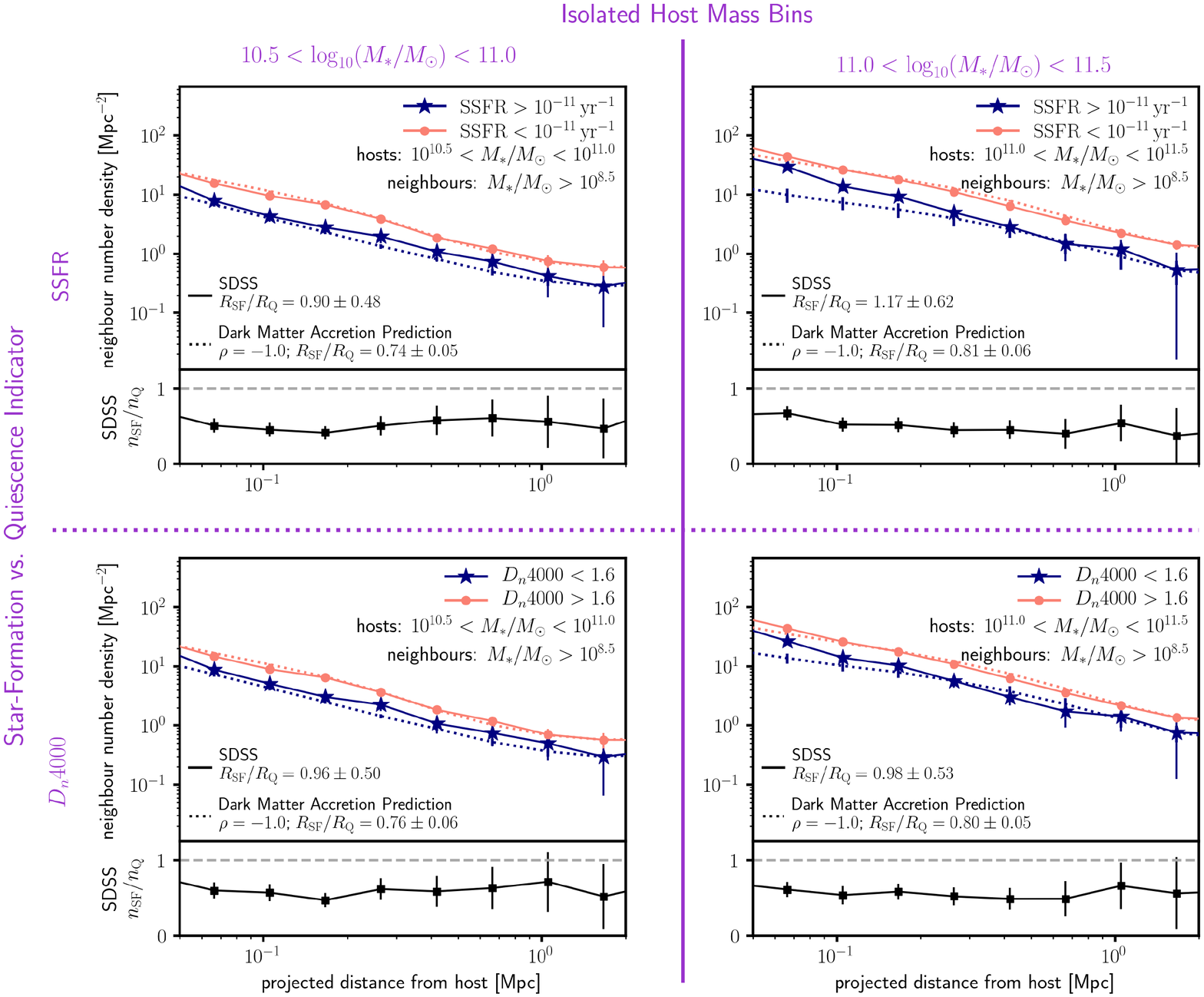}
    \caption{Same as Fig.~\ref{fig:results_9.0}, but with neighbours with $M_* > 10^{8.5} M_\odot$. We note that neighbours at these lower masses are not observable in SDSS throughout the isolated host redshift range. For isolated hosts with $10.5 < \log_{10}(M_*/M_\odot) < 11.0$, the SDSS observation limits for are $M_* > 10^{8.95} M_\odot$ at the median redshift $z=0.079$ and $M_* > 10^{9.36} M_\odot$ at the maximum redshift $z=0.123$. Similarly, for isolated hosts with $11.0 < \log_{10}(M_*/M_\odot) < 11.5$, the SDSS observation limits are $M_* > 10^{9.30} M_\odot$ at the median redshift $z=0.116$ and $M_* > 10^{10.4} M_\odot$ at the maximum redshift $z=0.183$.}
    \label{fig:results_8.5}
\end{figure*}

\begin{figure*}
    \centering
    \includegraphics[width=\textwidth]{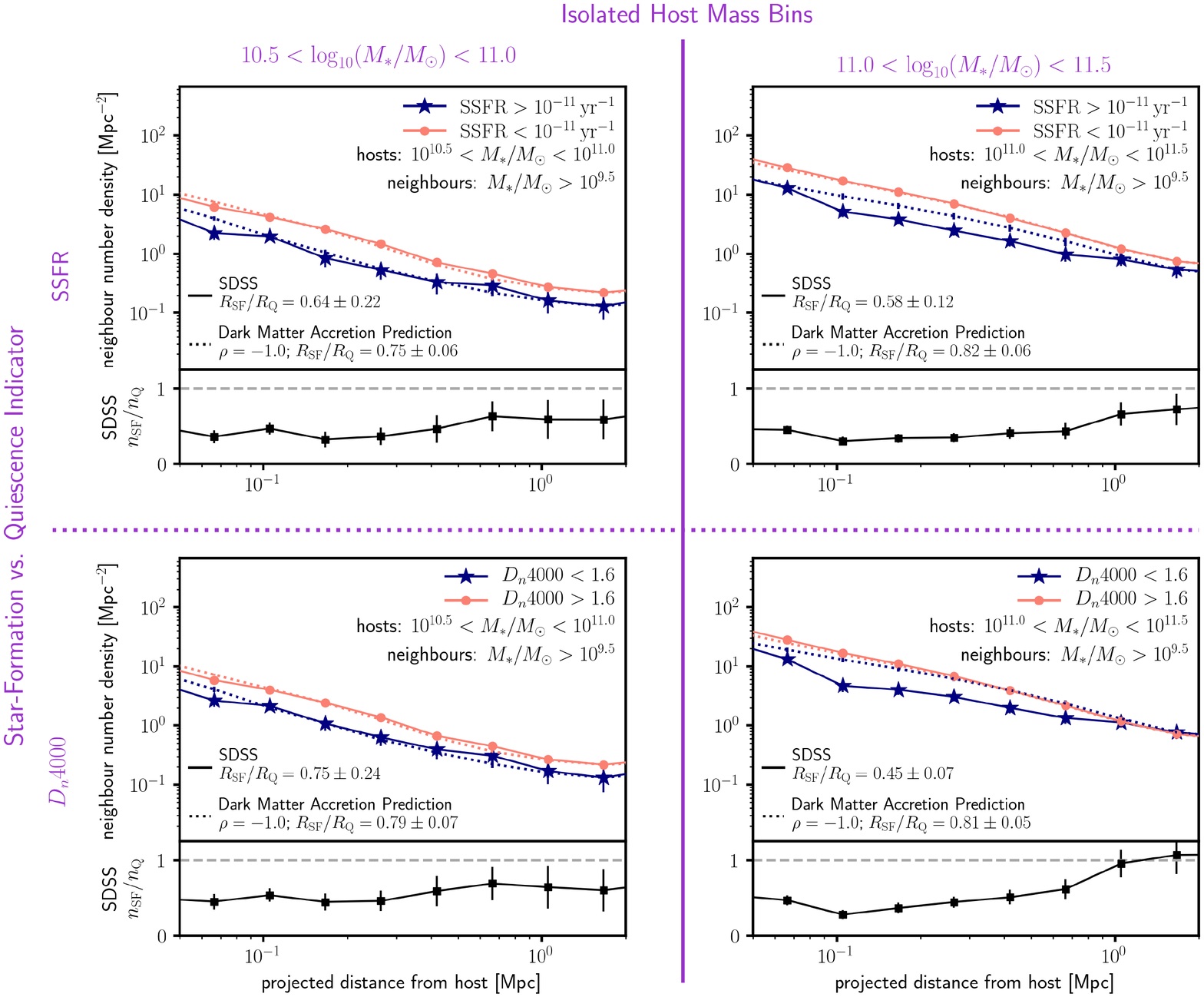}
    \caption{Same as Fig.~\ref{fig:results_9.0}, but with neighbours with $M_* > 10^{9.5} M_\odot$. These neighbour density distributions are also consistent with $\rho \leq 0.0$ with $\gtrsim 85$ per cent confidence.}
    \label{fig:results_9.5}
\end{figure*}

\begin{figure*}
    \centering
    \includegraphics[width=\textwidth]{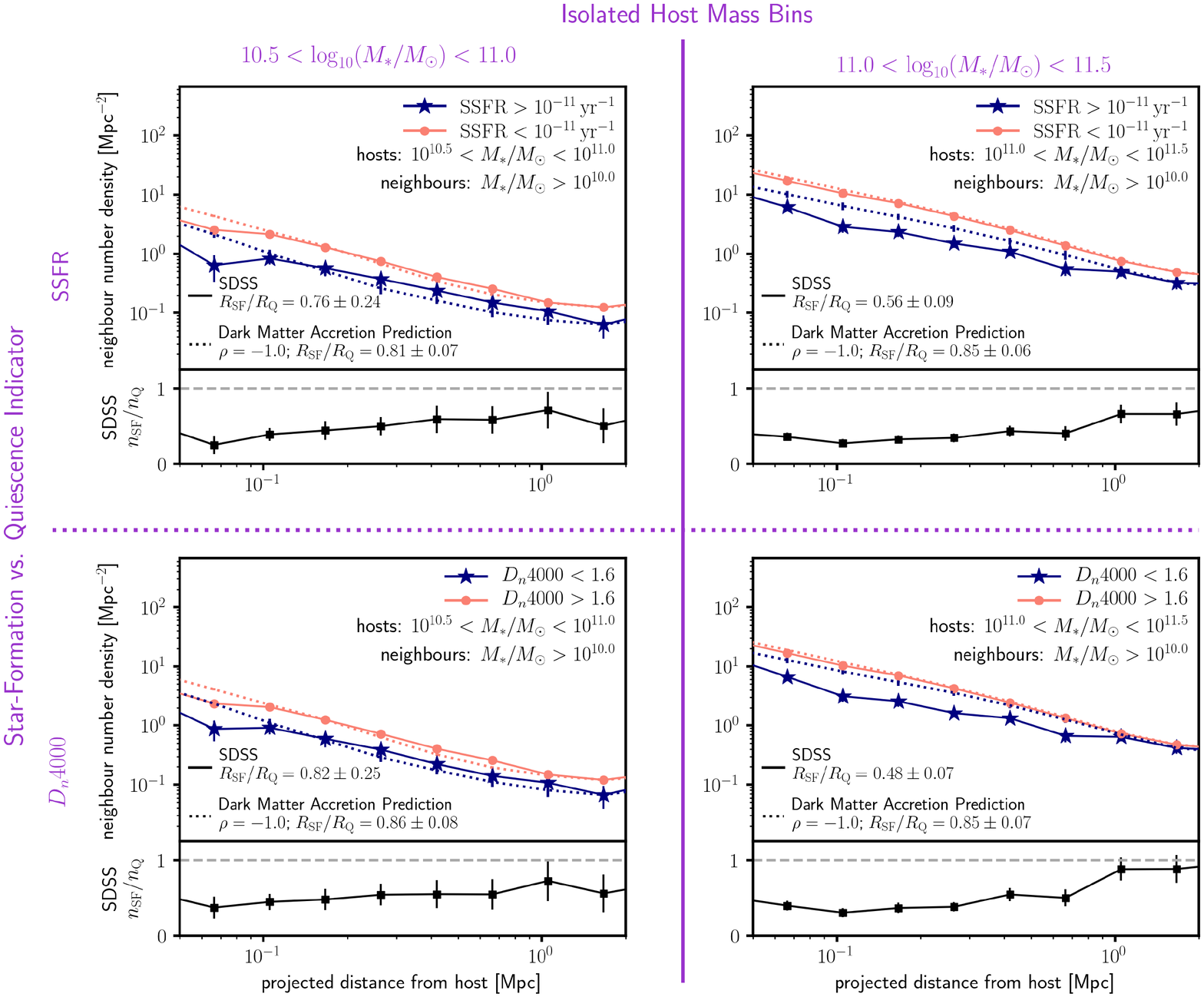}
    \caption{Same as Fig.~\ref{fig:results_9.0}, but with neighbours with $M_* > 10^{10.0} M_\odot$. These neighbour density distributions are also consistent with $\rho \leq 0.0$ with $\gtrsim 85$ per cent confidence.}
    \label{fig:results_10.0}
\end{figure*}

\label{lastpage}
\end{document}